\documentclass[aps,prf,amsmath,amssymb,amsfonts,groupedaddress,noshowpacs]{revtex4-1}
\usepackage{epsfig}
\usepackage{graphicx}
\usepackage{dcolumn}
\usepackage{bm}
\usepackage{amsmath}
\usepackage[normalem]{ulem}

\usepackage[titletoc,title]{appendix}

\usepackage{xcolor}

\begin{document}

\author{Itzhak Fouxon$^{1,2}$} \email{itzhak8@gmail.com} \author{Boris Rubinstein$^3$} \email{bru@stowers.org}
\author{Zhouyang Ge$^4$}\email{zhoge@mech.kth.se} \author{Luca Brandt$^4$}\email{luca@mech.kth.se}
\author{Alexander Leshansky$^1$}\email{lisha@tx.technion.ac.il}
\affiliation{$^1$ Department of Chemical Engineering, Technion, Haifa 32000, Israel}
\affiliation{$^2$ Department of Computational Science and Engineering, Yonsei University, Seoul 120-749, South Korea}
\affiliation{$^3$ Stowers Institute for Medical Research, 1000 E 50th st.,Kansas City, MO 64110, USA}
\affiliation{$^4$ Linn\'{e} FLOW Centre and SeRC (Swedish e-Science Research Centre), KTH Mechanics, SE-100 44 Stockholm, Sweden}
\begin{abstract}

The seminal Batchelor-Green's (BG) theory on the hydrodynamic interaction of two spherical particles of radii $a$ suspended in a viscous shear flow neglects the effect of the boundaries. In the present paper we study how a plane wall modifies this interaction. Using an integral equation for the surface traction we derive the expression for the particles' relative velocity as a sum of the BG's velocity and the term due to the presence of a wall at finite distance, $z_0$. Our calculation is not the perturbation theory of the BG solution, so the contribution due to the wall is not necessarily small. We indeed demonstrate that the presence of the wall is a singular perturbation, i.e., its effect cannot be neglected even at large distances. The distance at which the wall significantly alters the particles interaction scales as $z_0^{3/5}$. The phase portrait of the particles' relative motion is different from the BG theory, where there are two singly-connected regions of open and closed trajectories both of infinite volume. For finite $z_0$, besides the BG's domains of open and closed trajectories, there is a domain of closed (dancing) and open (swapping) trajectories that do not materialize in an unbounded shear flow. The width of this region grows as $1/z_0$ at smaller separations from the wall. Along the swapping trajectories, that have been previously observed numerically, the incoming particle is turning back after the encounter with the reference particle, rather than passing it by, as the BG theory anticipates. The region of dancing trajectories has infinite volume and is separated from a BG-type domain of closed trajectories that becomes compact due to presence of the wall. We found a one-parameter family of equilibrium states that were previously overlooked, whereas the pair of spheres flows as a whole without changing its configuration. These states are marginally stable and their perturbation yields a two-parameter family of the dancing trajectories, whereas the test particle is orbiting around a fixed point in a frame co-moving with the reference particle. We suggest that the phase portrait obtained at $z_0 \gg a$ is topologically stable and can be extended down to rather small $z_0$ of several particle diameters. We confirm this hypothesis by direct numerical simulations of the Navier-Stokes equations with $z_0=5a$.

\end{abstract}

\title{The effect of a wall on the interaction of two spheres in shear flow: \\ Batchelor-Green theory revisited}
\maketitle

\section{Introduction}

Small particles, droplets and bubbles are ubiquitously present in flowing fluids. When a suspended particle is transported by a viscous fluid, it modifies the flow around it. If another particle happens to be in the region of the modified flow, mutual hydrodynamic interactions between the particles will take place. The interactions are given implicitly by imposing boundary conditions on the flow that must hold simultaneously on the surfaces of all interacting particles \cite{hb,kim}. This setting is inconvenient for analyses, both theoretical and numerical. Thus there is no answer to even simplest questions, for instance whether there can be a non-trivial stationary configuration of particles that would flow as a whole due to the hydrodynamic interaction. Although these interactions somewhat resemble electrostatic interactions, there is no a hydrodynamic counterpart of Earnshaw's theorem \cite{pur} stating that such simple configurations are impossible. Here we provide an example of this possibility in the presence of a boundary and demonstrate that boundaries can have surprising and  non-trivial effects on hydrodynamic interactions.

\begin{figure}[t]
\includegraphics[width=0.4\textwidth]{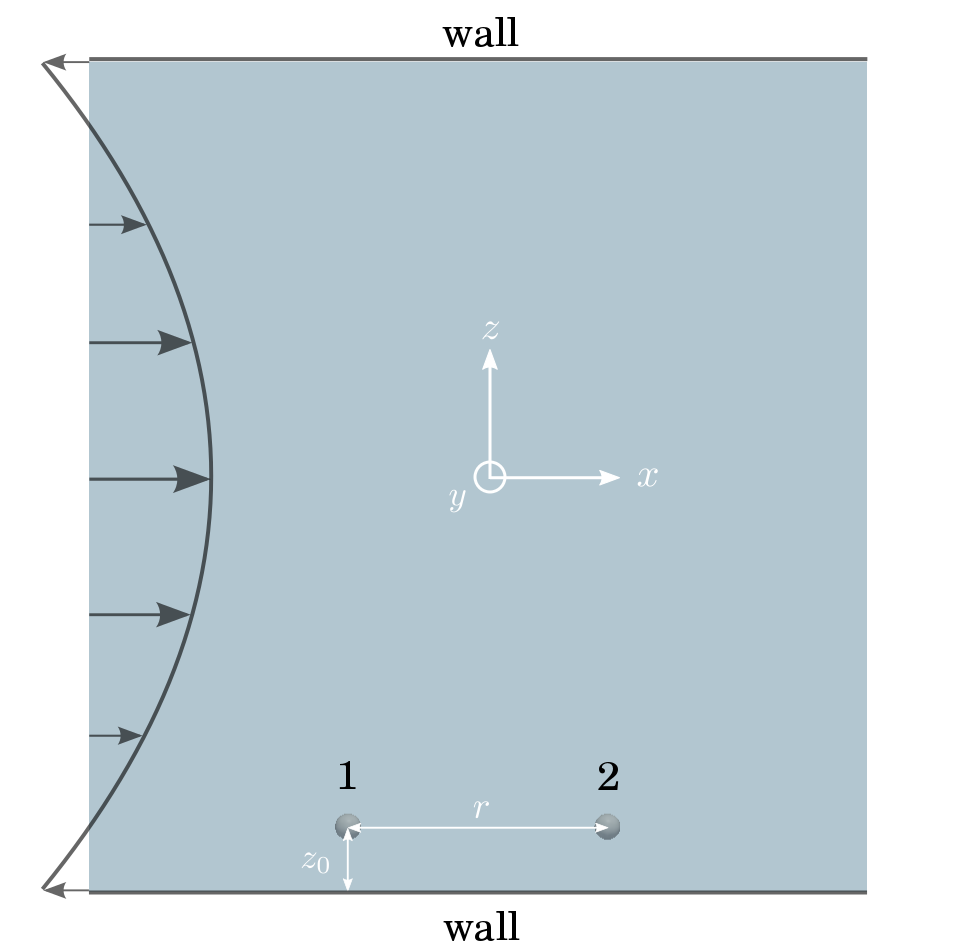}
\caption{Setup of a particle pair in the Poiseuille flow (depicted in a comoving reference frame). In this work we study the case $z_0 \gg a$, where $a$ is the particle radius, however does not necessarily require $z_0\gg r$. The upper wall is assumed to be much further away from the particle pair than $z_0$.
}
\label{fig:setup}
\end{figure}

The only well-studied case of hydrodynamic interactions of particles transported by non-uniform flow is the case of two particles in a time-independent low Reynolds number linear flow. This was studied in the seminal Batchelor-Green's  paper \cite{ujhd}, see also \cite{arp} for more details and an account of various contributions to the problem. If the particles' (and fluid) inertia can be entirely neglected, the vector between the particle centers obeys an autonomous first-order evolution equation, which gives its rate of change as a function of the instantaneous value. The use of the symmetries makes it possible to quantify the interaction by the two scalar functions of the distance, which have been tabulated \cite{ujhd}, see also \cite{kim}. This case presents no stationary configurations for the two particles. One of the main applications is the Poiseuille flow in the channel shown in Fig. \ref{fig:setup}. If both particles are far from the walls, $z_0 \gg a, r$, they can seemingly be considered as flowing in an unbounded shear flow and the analysis of \cite{ujhd,arp,lin} applies. The BG theory thus predicts that there are no possible stationary configurations of the particle pair. Here we demonstrate that the approximation of an unbounded flow overlooks such configurations and also other phenomena, which hold independently of how large $z_0$ is, cf. \cite{agto}. Thus, the presence of the wall exhibits a singular perturbation of the BG theory.

Recently, stationary configurations of particles transported in microfluidic channels attracted attention due to the possibility of flow-assisted microfabrication by using a combination of hydrodynamic and non-hydrodynamic (i.e., adhesive) interactions \cite{tab0,tabeling}. Under certain conditions, the particles can self-assemble into clusters of different morphology that flow with no change of inter-particle distances, see \cite{flow-assist} for detailed discussions. These micron-scale clusters can then be solidified and collected from the flow, and be potentially used for fabrication of functional metamaterials. To theoretically explain and subsequently predict the structure formation of the suspended particles observed in experiments, Shen \textit{et al}. \cite{tabeling} proposed a
model based on a dipolar asymptotic form of hydrodynamic interaction. Notice that the dipolar form only holds at large particle separations (it was derived in detail using the fundamental solution for the channel flow \cite{LironMochon}, see \cite{2017} and also references therein). One reason for the apparent applicability of this description, despite the particles were \textit{close to} each other, is likely the dominance of the adhesive radial forces between particles at close proximity, such that any hydrodynamic interaction producing a non-zero tangential velocity component would yield a similar cluster formation dynamics (see \cite{flow-assist} for further evidence). In contrast, a consistent predictive theory of hydrodynamic interactions should hold irrespective of the presence of adhesive forces, and allow for analysis of interaction of flowing particles at close proximity and near the wall, as in experimental setup \cite{tabeling}. The present paper is a step toward this theory.

There are two differences between the channel flow and the unbounded shear flow considered in the BG theory  \cite{lin,ujhd}. The velocity profile of the channel flow is quadratic \cite{quadratic} in the coordinate rather than linear. This difference is often irrelevant when the interacting particles are located much closer to one of the channel walls, so that the flow can be closely approximated by the linear shear flow. This is the case we consider in the present paper. Another difference, is that the no-slip rigid wall is always at a finite distance and it interacts with the flowing particles.

\begin{figure}[t]
\includegraphics[width=0.4\textwidth]{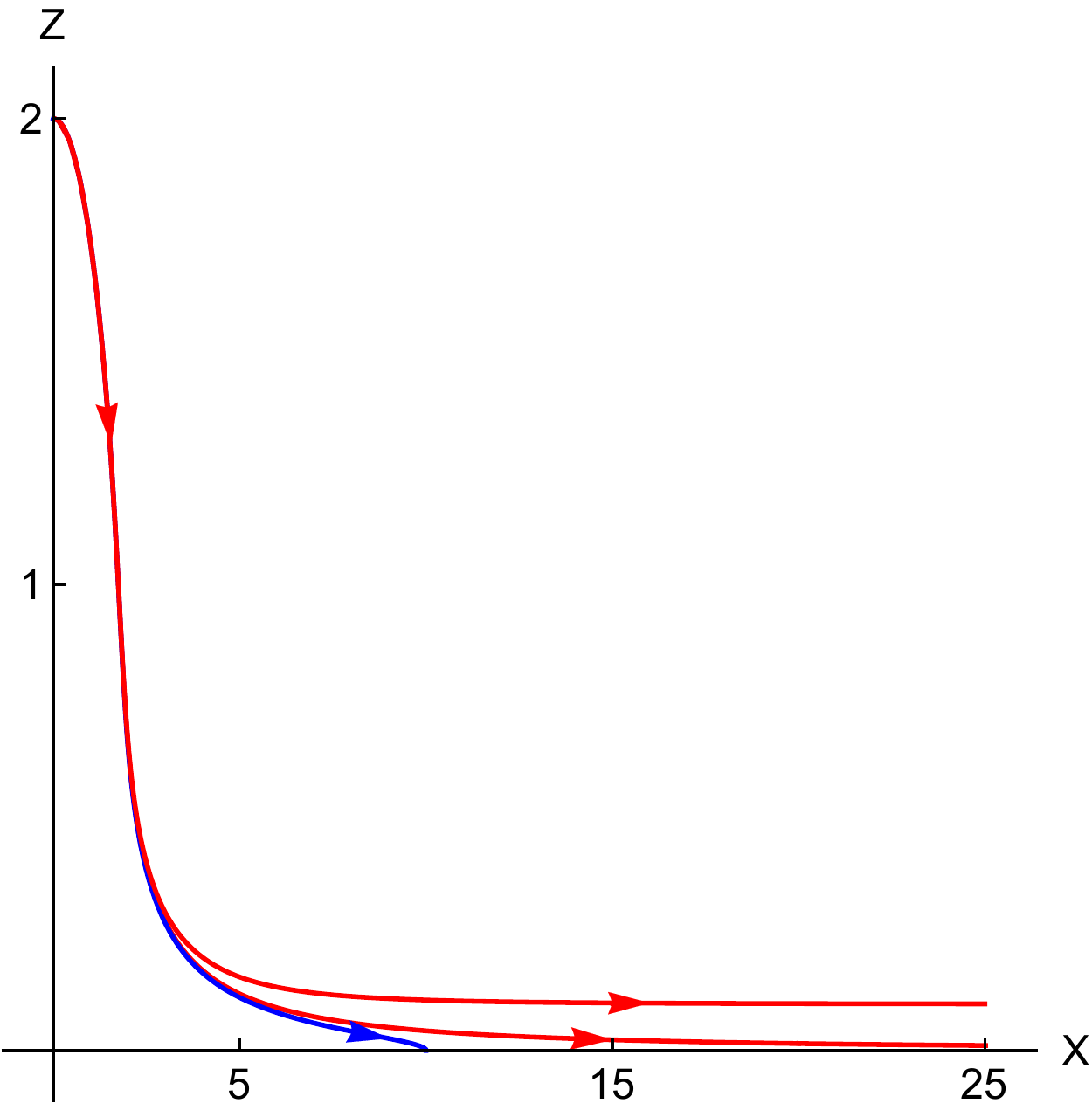}
\caption{Phase portrait of the trajectories in the symmetry $xz$-plane, within the BG approximation of $z_0=\infty$. All lengths are scaled with the particle radius $a$.
Throughout the paper the reference sphere is at the origin and the trajectories of the second sphere are shown. Due to the fore-and-aft and top-down symmetries only one quadrant is depicted. The two types of the trajectories -- closed (blue) and open (red), are separated by the separatrix, the open trajectory that asymptotically approaches the $x$-axis \cite{lin,ujhd}. For two spheres at the same vertical line, the maximal separation for closed trajectories is of order of $10^{-5}$, see \cite{arp}. As a result, at this resolution, the trajectories are indistinguishable when approaching the $z$-axis. This includes the shown open trajectory that crosses the $z$-axis slightly above the closed trajectories. The time-period of revolution along the shown closed trajectory is more than $700$ (here and thereafter the time units of inverse shear rate $\dot{\gamma}^{-1}$ are used).}
\label{fig:SeparatrixBG}
\end{figure}

We first consider the evolution of the inter-particle distance when the effect of the wall is neglected and the BG theory applies. It is useful to consider the three-dimensional phase space spanned by all possible distances $\bm r$ between the spheres' centers where one of them is at the origin. The distance $\bm r(t)$ between the spheres' centers obeys an autonomous evolution equation which means that there is a well-defined phase space flow $\bm V^0(\bm r)$ such that $\dot{\bm r}=\bm V^0(\bm r(t))$ and a unique trajectory passes through each point. This is the consequence of neglecting particles' and fluid inertia and the translational invariance due to which the shear resistance matrix depends on $\bm r$ only, cf. \cite{sh} (translation in a linear flow changes the flow by a constant vector, irrelevant by Galilean invariance). The flow $\bm V^0(\bm r)$ does not vanish anywhere so that there are no steady configurations. The absence of critical points with $\bm V^0(\bm r)=0$  implies a simple structure of the phase space. This can be most readily observed in the symmetry plane formed by the horizontal flow direction $x$ and the vertical velocity gradient direction $z$ (see Fig.~\ref{fig:SeparatrixBG}). The trajectories that belong to the plane never leave it, $V_y(y=0)=0$, and can be considered separately. There is a simple dichotomy of the trajectories:  closed trajectories crossing the $x$-axis and open trajectories that do not cross the $x$-axis. The open trajectories describe the faster particle overtaking the slower one. The particles return to their original vertical positions following the hydrodynamic encounter and there is fore-and-aft symmetry of the phase portrait. In contrast, the trajectories that cross the $x$-axis are closed, corresponding to a bound pair of spheres orbiting around each other. Open and closed trajectories are separated by the separatrix that touches the $x$-axis asymptotically at large distances \cite{lin}. Rotation of this separatrix around the $z$-axis creates an axisymmetric surface that separates the regions of open and closed trajectories in space (it is not readily evident how this axial symmetry could be guessed \textit{a priori} without writing down the equations). The region of closed trajectories has an infinite volume, which presents difficulties in, e.g., calculation of the effective viscosity of a dilute hard-sphere suspension at the quadratic order in concentration \cite{bgst}.
\begin{figure}[t]
\includegraphics[width=0.4\textwidth]{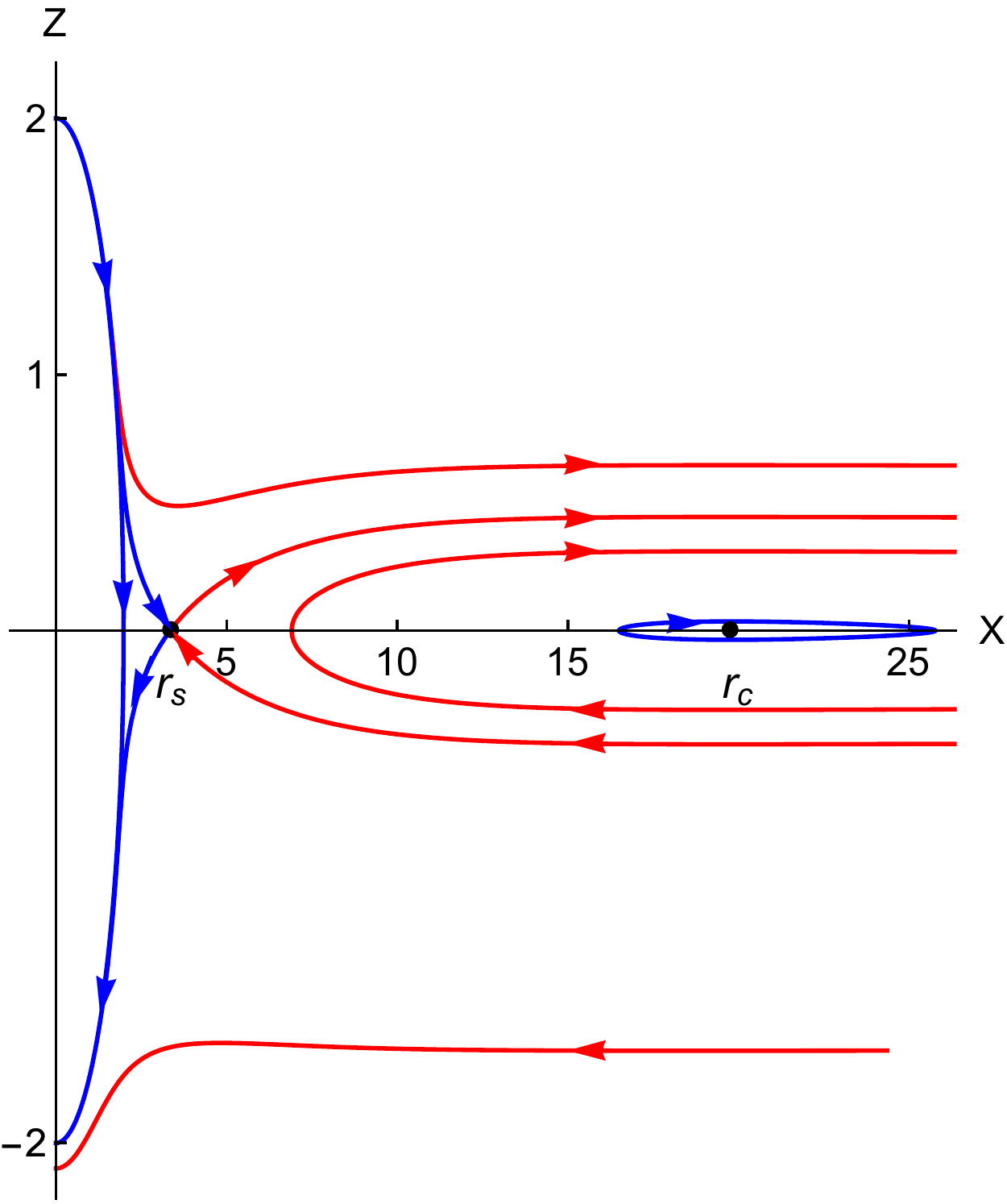}
\caption{The phase portrait in the symmetry $xz$-plane at finite $z_0=5$. The fore-and-aft and top-down symmetries survive the wall perturbation in the leading order. The phase portrait exhibits two critical (equilibrium) points: the saddle (hyperbolic) point $r_s$ and the neutral equilibrium (elliptic) point $r_c$, representing a completely different topology from the BG theory in Fig.~\ref{fig:SeparatrixBG}. As $z_0 \to \infty$, the topology of the phase portrait is preserved, while the critical points are being shifted to infinity. }
\label{fig:Schematics}
\end{figure}

\begin{figure}[t]
\begin{center}
\begin{tabular}{ccc}
\includegraphics[width=0.35\textwidth]{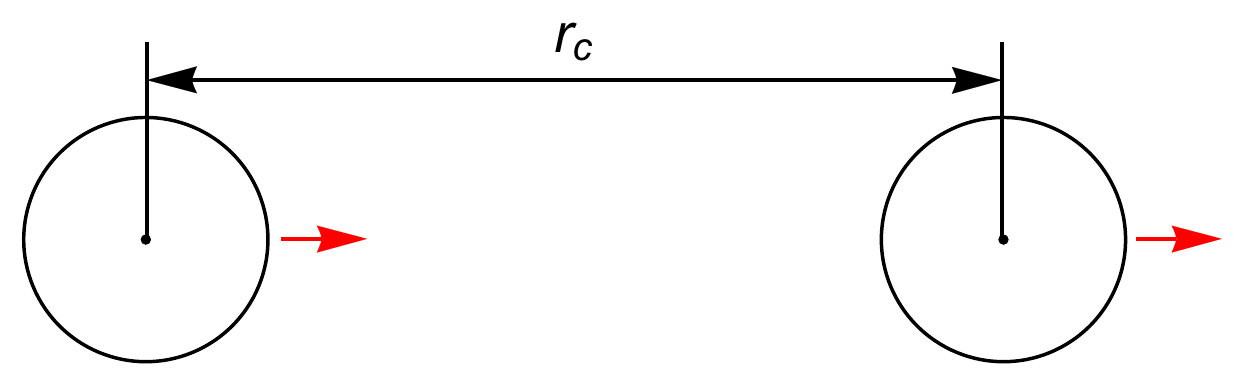} &
\hskip1cm&
\includegraphics[width=0.35\textwidth]{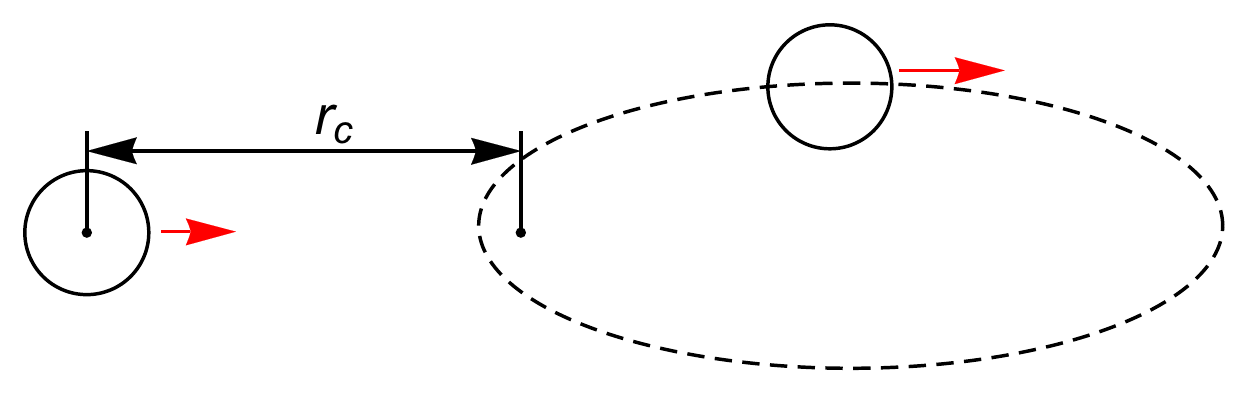} \\
(a) && (b)
\end{tabular}
\end{center}
\caption{(a) There is a unique value of the stream-wise separation distance between two particles, $x=r_c$, flowing along the same streamline of the Poiseuille flow, for which they flow steadily without changing their configuration. The stability of such motion is marginal, cf. (b). The value $r_c=4 z_0$ is confirmed in the numerical simulations of the Navier-Stokes equations for $z_0=5$.
(b) Trajectories that pass through points around the stationary point $x=r_c$, $z=0$, crossing $x$-axis at distance larger than $2\sqrt{2}z_0$, exhibit a peculiar dancing dynamics. In the coordinate frame co-moving with the trailing (left) particle, the leading (right) sphere follows an elongated closed orbit around $(r_c, 0)$. Similar trajectories hold outside the symmetry plane.}
\label{Draw1}
\end{figure}

We demonstrate here that when a distant wall is considered, the evolution of $\bm r$ remains autonomous in the leading approximation, $\dot{\bm r}=\bm V(\bm r(t))$. Thus at any finite $z_0\gg a$  we can still examine the phase portrait, which is however qualitatively different from that in Fig.~\ref{fig:SeparatrixBG}. Our calculation is not a perturbation theory of the BG solution as we do not assume $\bm V\approx \bm V^0$, so the disturbed phase space flow $\bm V(\bm r)$ is significantly different from $\bm V^0$. The change in topology occurs because at finite $z_0$ there exist critical points at which $\bm V(\bm r)$ vanishes, see the phase portrait in the symmetry plane in Fig.~\ref{fig:Schematics}. The saddle (hyperbolic) point $r_s$, the closer of the two critical points to the origin, is unstable. The other neutral equilibrium (elliptic) point $r_c$ corresponds to a marginally stable configuration, where the pair flows without changing its inter-particle distance and orientation, see Fig.~\ref{Draw1}(a). Not too large deviations from this state result in the sphere orbiting around this elliptic point. These dancing trajectories would have rather unusual appearance when considered in the laboratory frame: while one sphere travels downstream, the other sphere revolves around a point co-moving in space with the first sphere, see Fig.~\ref{Draw1}(b). The phase plane at $x>0$ is characterized by two disconnected regions of closed trajectories shown in Fig.~\ref{fig:Schematics} by the blue curves. The region to the left of
the separatrix (blue) that crosses $x=r_s$ resembles the BG's closed trajectories. The trajectories around the elliptic point $(r_c, 0)$ are solely due to the presence of the wall. The lowest red curve is an open trajectory similar to the BG's: after the encounter the vertical separation is restored to its initial value. In contrast to that, along the (red) trajectories circumventing the elliptic point, the vertical separation of the particles reverses sign after the encounter. We call these ``swapping trajectories", since they seem to correspond to the numerical findings of \cite{agto} at $z_0\sim a$, where trajectories with particles swapping their vertical positions after the encounter in a channel flow were reported (careful consideration of the Figure presented in \cite{agto} reveals slight changes of the vertical coordinates which seem to be a higher order effect than that considered here). The phase portrait for the evolution of inter-particle distance along the swapping trajectory obtained in \cite{agto} numerically, agrees with that predicted here theoretically. To prove that the sign-reversal of the vertical separation predicted here implies swapping, it has to be shown that the center of mass of the pair is not displaced vertically as a result of the encounter. We leave the rigorous proof for future work, focusing here on the evolution of inter-particle distance only. Thus, the use of the term ``swapping trajectories" here, strictly speaking, refers to open trajectories with sign-reversal of the vertical separation following the encounter.

\begin{figure}[t]
\includegraphics[width=0.4\textwidth]{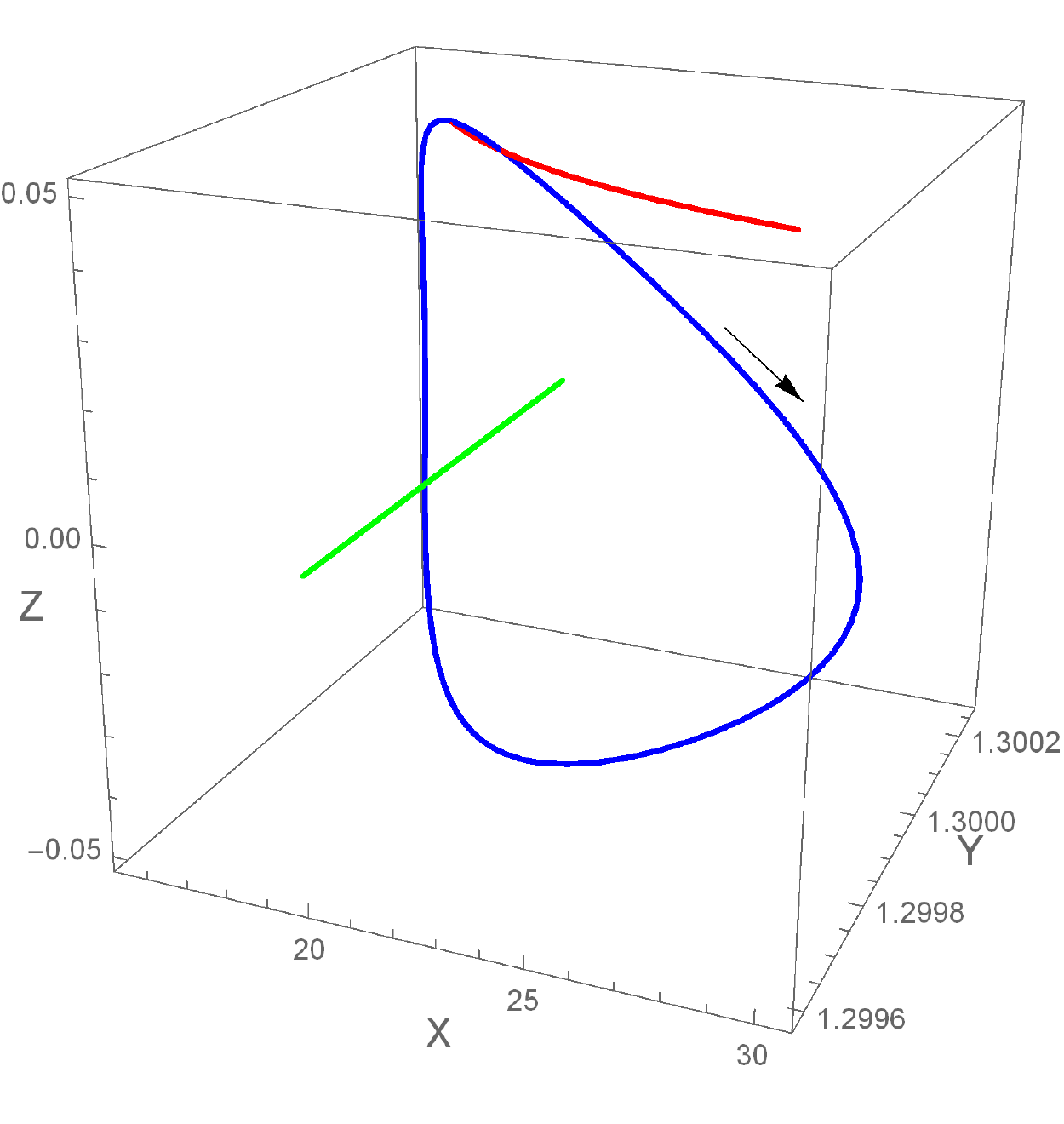}
\caption{The blue line depicts the closed trajectory that forms a loop around the critical circle of radius $r_c$ (whose segment is shown by the green line) for $z_0=5$.
The trajectory can be shrunk to a (necessarily critical) point by continuously changing the initial conditions. The red line shows the BG trajectory
that starts from the same initial condition as the blue line. The period of revolution along the closed trajectory is $1165$.}
\label{Fig8b_CircleRc}
\end{figure}

\begin{figure}[t]
\includegraphics[width=0.4\textwidth]{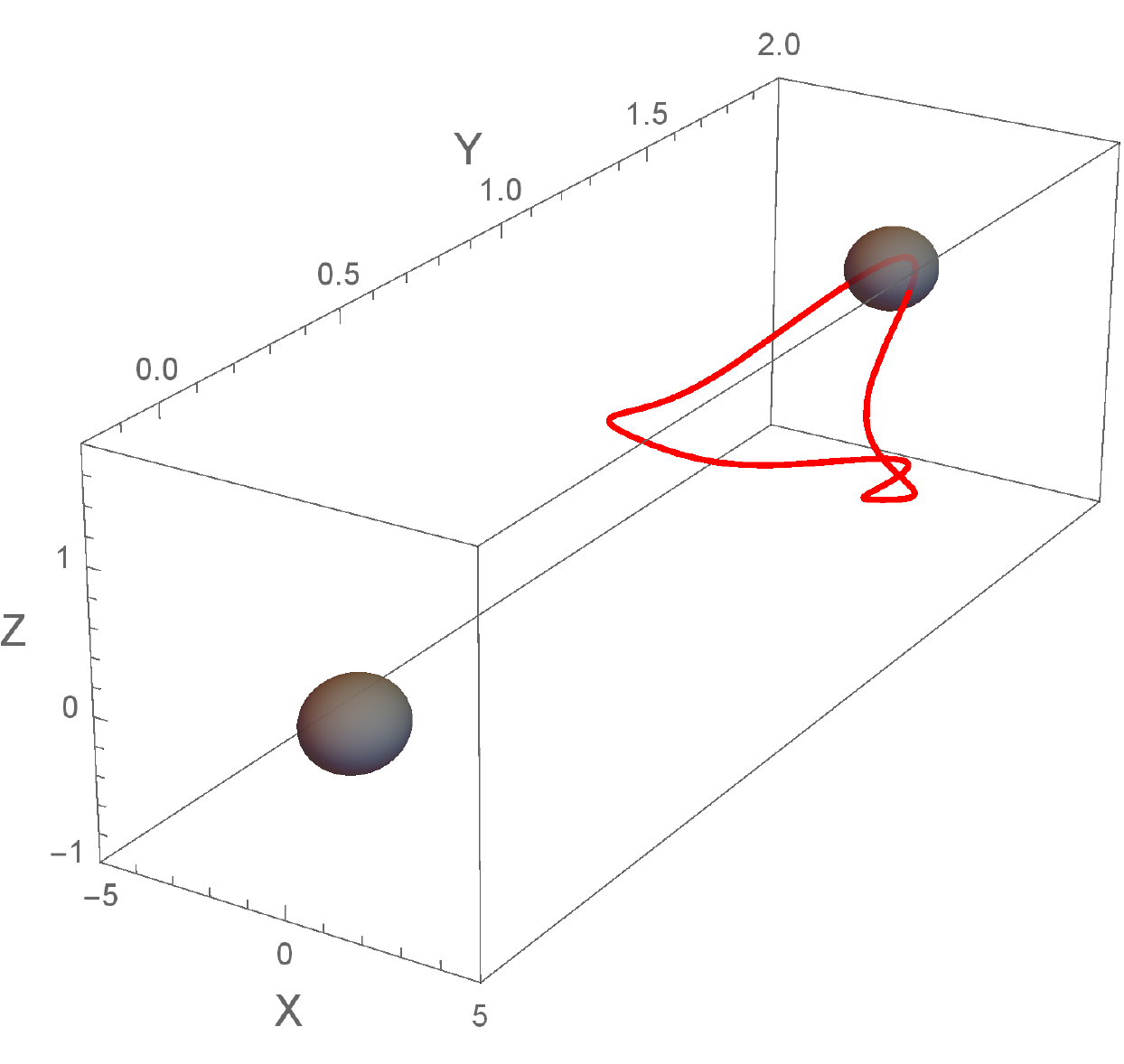}
\caption{Three-dimensional BG trajectory (red curve) of the sphere in the frame co-moving with the reference sphere. The closed trajectory has a geometrical center on the $y$-axis. In contrast to the trajectory in Fig.~\ref{Fig8b_CircleRc}, this curve cannot be shrunk to a point by a continuous change of the initial conditions. That point would have to be critical and the BG phase space does not admit those. For this trajectory, a distant wall is only a regular small perturbation. 
The period of revolution is $120$.}
\label{FigX_BG_closed_trajectory1}
\end{figure}

The three-dimensional trajectories are more complex. The circle of radius $r_c$ around the $z$-axis provides the critical curve with $\bm V(\bm r)=0$. The configurations with $\bm r$ on that circle are stationary, so that for instance there is a stationary pair where only the $y$-coordinates of the particles are different. Displacements from these stationary configurations result in closed trajectories that loop around the critical circle, see Fig.~\ref{Fig8b_CircleRc}. In contrast with the symmetry plane, where the BG trajectories display no behavior similar to dancing, some of the three-dimensional BG trajectories do look rather similar (notice that it was not stressed in the original work or in \cite{arp}), see Fig.~\ref{FigX_BG_closed_trajectory1} for comparison.

We emphasize the topological difference between the phase portrait in Fig.~\ref{fig:Schematics} and that of the BG theory in Fig.~\ref{fig:SeparatrixBG}. There are two disconnected regions of closed trajectories. In one region the particles orbit each other, similarly to $z_0=\infty$ approximation, however the volume of this region is finite. The other region contains dancing closed trajectories and at large $x^2+y^2$ it is bounded by the surface of revolution $|z|\propto z_0(x^2+y^2)^{-3/2}$. This is similar to the BG bounding surface, $|z|\propto (x^2+y^2)^{-3/2}$, however boosted by the large $z_0$ factor. In both cases the volume of the phase space domain containing closed trajectories diverges, so the divergences in the second order in particle concentration stress calculations of \cite{bgst} are not regularized by the wall. At finite $z_0$ the two regions of closed trajectories are separated by a region of a new type of open swapping trajectories that, in contrast to the BG theory, cross the $x$-axis. Then the top-down symmetry, which holds remarkably in the presence of the wall, implies that for open trajectories that cross the $x$-axis, the vertical component of the inter-particle distance reverses its sign, as in numerically observed swapping trajectories \cite{agto}. At least some features of the presented topology, derived theoretically at $z_0\gg a$, work accurately down to $z_0=5a$, as demonstrated by our in-house numerical simulations of the Navier-Stokes equations.

In the next Section we present the detailed derivation of the evolution equation for the inter-particle distance in the wall-bounded shear flow. Sec.~\ref{singular} demonstrates why the wall presents a singular perturbation of the BG theory. In Sec.~\ref{infinitely} we review the BG trajectories that serve as the reference point of our study. We present the results of the numerical solutions of the derived evolution equation in Sec.~\ref{finitely}. Sec. \ref{dancing-swapping} presents full solution of equation of motion in the dancing-swapping region. Section \ref{dns} presents the confirmation of the theory by direct numerical simulations of the motion of a pair of spheres in the Poiseuille flow. In the last Section we conclude our results, discuss the applicability of the BG theory and formulate some open questions.

\section{Evolution of the distance between two particles transported by a shear flow near wall} \label{interpair}

In this Section we derive the autonomous evolution equation for the distance $\bm r(t)$ between two spheres transported by the Poiseuille flow. We make the simplifying assumption that both spheres are much closer to one of the bounding walls than the other. Thus the particles are effectively transported by the shear flow and not the parabolic velocity profile. The hydrodynamic interaction of particles transported by an unbounded shear flow are well-studied and their velocities
$\bm V^0_{\alpha},\ \alpha=1,2$ were considered in \cite{ujhd}. This analysis serves as a starting point of our study. We also assume that the distance to the wall is much larger than the particles' radii. We derive the particles' relative velocity as a sum of $\bm V^0_{\alpha}$ and the correction velocity $\delta \bm V_{\alpha}$. The correction velocity is not necessarily smaller than $\bm V^0$, as our solution is not a perturbation around the solution for an unbounded shear flow.

\subsection{Direct approach} \label{direct}

We set the problem and consider its formulation using the flow for infinitely separated walls as a reference.
The problem of two spheres driven by the Poiseuille flow is described by,
\begin{eqnarray}&&\!\!\!\!\!\!\!\!
\nabla p\!=\!\eta\nabla^2\bm u,\ \ \nabla\!\cdot\!\bm u\!=\!0,\ \ \ \ \bm u(z\!=\!0)\!=\!\bm u(z\!=\!h)\!=\!0,\ \ 
u_x(\infty)\!=\!\frac{z(z\!-\!h)\nabla_x p^0}{2\eta},\ \ \bm u(S_{\alpha})\!=\!\bm V_{\alpha}\!+\!\bm \Omega_{\alpha}\!\times\! (\bm x\!-\!\bm x_{\alpha}),
\label{ds}
\end{eqnarray}
where $\alpha=1, 2$ are the indices of the spheres, $x_{\alpha}$ are the coordinates of the centers and $S_{\alpha}$ is the surface of the $\alpha-$th sphere. We designate the flow by $\bm u$, and the translational and rotational velocities of the spheres by $\bm V_{\alpha}$ and $\bm \Omega_{\alpha}$. The constant pressure gradient $\nabla p^0=-|\nabla_x p^0| \bm{\hat x}$ drives the flow in the positive $x-$direction, $\eta$ is the fluid viscosity, $z$ is the vertical coordinate and $h$ the channel height. We assume that the spheres have equal radii $a$, although most of the calculations below can be done without this assumption.
We will use below $a$ as the unit of length so that the radii are 1.
We assume that the particle inertia is negligible so that the values of $\bm V_{\alpha}$ and $\bm \Omega_{\alpha}$ are determined from the conditions that the total force and torque from the fluid on either particle is zero,
\begin{eqnarray}&&\!\!\!\!\!\!\!\!\!\!\!\!\!
\int_{S_{\alpha}} \bm t  dS=0,\ \ \int_{S_{\alpha}} (\bm x-\bm x_{\alpha})\times \bm t  dS=0, \label{inertless}
\end{eqnarray}
where we have introduced the surface traction $\bm t$, which can be written via the stress tensor $\sigma_{ik}$ as,
\begin{eqnarray}&&\!\!\!\!\!\!\!\!\!\!\!\!\!
t_i(\bm x)\!=\!\frac{\sigma_{ik}(\bm x\!-\!\bm x_{\alpha})_k}{a},\ \ \sigma_{ik}\!=\!-p\delta_{ik}\!+\!\eta(\nabla_iu_k\!+\!\nabla_ku_i),
\end{eqnarray}
where $\bm x$ belongs to $S_{\alpha}$. We observe that if the spheres are much closer to the wall at $z=0$ than at $z=h$, then we can use different boundary conditions in Eq.~(\ref{ds}),
\begin{eqnarray}&&\!\!\!\!\!\!\!\!\!\!\!\!\!
\bm u(z=0)=0,\ \ \bm u(\infty)\!=\!\dot{\gamma}z\bm{\hat x}, \ \ \dot{\gamma}\!=\!\frac{h|\nabla_x p^0| }{2\eta},
\end{eqnarray}
where we introduced the effective shear rate $\dot{\gamma}$ in terms of the parameters defining the Poiseuille flow. The boundary conditions at $S_{\alpha}$ are unchanged. Without the boundary condition at $z=0$ we reduce to the problem of motion of two spheres in an unbounded shear flow considered in \cite{ujhd}. We designate all quantities
of this problem by the superscript zero. Thus $\bm u^0$ is the unbounded shear flow for the two spheres with translational and rotational velocities $\bm V^0_{\alpha}$ and $\bm \Omega^0_{\alpha}$, which obey Eq.~(\ref{inertless}) with $\bm t=\bm t^0$. We look for the solution as superposition of $\bm u^0$ and the flow perturbation $\delta \bm u$ (where the use of $\delta$ does not imply smallness of $\delta \bm u$). We thus have,
\begin{eqnarray}&&\!\!\!\!\!\!\!\!
\nabla \delta p\!=\!\eta\nabla^2\delta\bm u,\ \ \nabla\cdot\delta\bm u\!=\!0,\ \ \ \ \delta\bm u(z\!=\!0)\!=\!-\bm u^0(z\!=\!0),\ \ 
\delta \bm u(\infty)\!=\!0,\ \ \delta \bm u(S_{\alpha})\!=\!\delta \bm V_{\alpha}\!+\!\delta\bm \Omega_{\alpha}\!\times\! (\bm x\!-\!\bm x_{\alpha}),
\label{ds1}
\end{eqnarray}
where we introduced deviations of the velocities and of the surface traction from their values in an infinite domain,
\begin{eqnarray}&&\!\!\!\!\!\!\!\!\!\!\!
\delta\bm V_{\alpha}\!=\!\bm V_{\alpha}\!-\!\bm V^0_{\alpha},\ \ \delta\bm \Omega_{\alpha}\!=\!\bm \Omega_{\alpha}\!-\!\bm \Omega^0_{\alpha},\ \ \delta \bm t\!=\!\bm t\!-\!\bm t^0.
\end{eqnarray}
The deviations of the velocities are fixed by the condition that the deviation from the surface traction obeys Eqs.~(\ref{inertless}) with $\delta \bm t$ instead of $\bm t$. We notice that the flow $\bm u^0(z=0)$ in Eq.~(\ref{ds1}) is induced by the spheres, since the unperturbed flow vanishes at $z=0$. Thus $\bm u^0(z=0)$ vanishes at infinity as necessary for consistency of the boundary conditions at the plane and at infinity. For the distant wall the flow $\bm u^0(z\!=\!0)$ can be simplified. This flow obeys the integral representation (see the derivation in Appendix \ref{shear}),
\begin{eqnarray}&&\!\!\!\!\!\!\!\!\!\!
u^0_i(\bm x)\!=\!\dot{\gamma}\delta_{i1}z \!-\!\sum_{\alpha}\int_{S_{\alpha}}\!\!\!\frac{Y_{il}(\bm x-\bm x')t^0_{l}(\bm x')dS}{8\pi \eta};\ \  Y_{il}(\bm r)=\frac{\delta_{il}}{r}+\frac{r_ir_l}{r^3},\ \ \bm r=\bm x-\bm x', \label{inter}
\end{eqnarray}
where $Y_{il}$ is the Oseen tensor or the Green's function of the Stokes flow in an unbounded fluid \cite{kim}. If the vertical positions $z_{\alpha}$ of  the centers of both spheres are much larger than their radii, $a$, then the asymptotic expansion of $\bm u^0(z\!=\!0)$ in $a/z_{\alpha}$ is obtained by Taylor expansion of $Y_{il}(\bm x-\bm x')$ in Eq.~(\ref{inter}) near $\bm x'=\bm x_{\alpha}$. Using the condition of zero force we find that, at the leading order,
\begin{eqnarray}&&\!\!\!\!\!\!\!\!\!\!
u^0_i(z=0)\!\approx \!\frac{1}{8\pi \eta}\frac{\partial}{\partial x_m}\sum_{\alpha} Y_{il}(\bm x-\bm x_{\alpha})S^{\alpha}_{lm}|_{z=0};\ \ \ S^{\alpha}_{lm}\!\equiv \!\!\int_{S_{\alpha}}\!\!\!\!\left(\!(\bm x\!-\!\bm x_{\alpha})_m t^0_l(\bm x)\!-\!\frac{\delta_{ml}(\bm x\!-\!\bm x_{\alpha})_pt^0_p(\bm x)}{3}\!\right)\!dS,
\label{far}
\end{eqnarray}
where the traceless tensor $S^{\alpha}_{lm}$ is \cite{ujhd} the force dipole strength of sphere $\alpha$. The $\delta_{lm}$ term  can be added since $\nabla_lY_{il}=0$. The force dipole strengths obey a general form derived in \cite{ujhd}. We have $S^{1}_{lm}=S^{2}_{lm}=S_{lm}$ with
\begin{eqnarray}&&\!\!\!\!\!\!\!\!\!\!\!
\frac{3S_{lm}(\bm r)}{10 \pi \eta a^3\dot{\gamma}}\!=\!\left(\delta_{lx}\delta_{mz}\!+\!\delta_{mx}\delta_{lz}\right)(1\!+\!K)
\!+\!\left(\frac{r_l\left(x\delta_{mz}\!+\!z\delta_{mx}\right)\!+\!r_m\left(x\delta_{lz}\!+\!z\delta_{lx}\right)}{r^2}\!-\!\frac{4xz\delta_{lm}}{3r^2}\right)L
\!+\!\frac{2xz}{r^2}\left(\frac{r_lr_m}{r^2}\!-\!\frac{\delta_{lm}}{3}\right) M,\label{dipole}
\end{eqnarray}
where the scalar functions $K$, $L$ and $M$ depend on the inter-particle distance $r/a$ only (we omitted the prime in the notation of \cite{ujhd}, as the spheres have identical radii in our case). These functions can be completely found only numerically and are considered below as given. We can use Eq.~(\ref{far}) instead of the boundary condition at $z=0$ in Eq.~(\ref{ds1}). The first reflection \cite{hb} gives the leading order approximation for $\delta\bm V_{\alpha}$ as in the Lorentz solution for a sphere in the presence of a distant wall \cite{hb}. The compact expansion can be found below from integral representations.

\subsection{Integral equation for velocities}

Here we derive the integral equation that determines the particle velocities. For future generalization to the case where the distances from the spheres to both walls are comparable we perform the derivation starting from the full formulation given by Eq.~(\ref{ds}).
We use the integral representation of the flow derived in \cite{2017},
\begin{eqnarray}&&\!\!\!\!\!\!\!\!\!
u_i(\bm x)\!=\!\frac{\delta_{ix}z(z\!-\!h)\nabla_x p^0}{2\eta}\!-\!\sum_{\alpha}\int_{S_{\alpha}}\!\!\!\frac{S_{il}(\bm x, \bm x')t_{l}(\bm x')dS'}{8\pi \eta},\label{inrepmany}
\end{eqnarray}
where we exploited the symmetry \cite{ps} of Green's function $S_{il}(\bm x, \bm x')=S_{li}(\bm x', \bm x)$. This function is defined by,
\begin{eqnarray}&&\!\!\!\!\!\!\!\!\!\!
\bm u^S(\bm x)=\frac{1}{8\pi \eta}S_{ik}(\bm x, \bm x_0)g_k, \label{vl}
\end{eqnarray}
where $\bm u^S$ is the Stokeslet flow caused by a point force acting between two parallel plates,
\begin{eqnarray}&&
-\nabla p^S+\eta \nabla^2 \bm u^S+\bm g\delta(\bm x-\bm x_0)=0,\ \ \nabla\cdot\bm u^S=0, \ \  \bm u^S(z=0)=\bm u^S(z=h)=0,\ \ \bm u^S(x^2+y^2\to\infty)=0.\label{stokes1}
\end{eqnarray}
The function $S_{ik}$ is independent of $\bm g$ and it was derived in \cite{LironMochon}. We study the velocities $\bm V_{\alpha}$ using the integral equation for the surface traction $\bm t(\bm x)$ obtained by taking
$\bm x$ in Eq.~(\ref{inrepmany}) to the surface of one of the spheres; this gives
\begin{eqnarray}&&\!\!\!\!\!\!\!\!\!\!\!\!\!\!\!\!
(V_{\alpha})_i+(\bm \Omega_{\alpha}\times (\bm x-\bm x_{\alpha}))_i\!=\!\frac{\delta_{ix}z(z\!-\!h)\nabla_x p^0}{2\eta}
-\!\sum_{\alpha'}\int_{S_{\alpha'}}\!\!\!\frac{S_{il}(\bm x, \bm x')t_{l}(\bm x')dS'}{8\pi \eta}, \label{intr1}
\end{eqnarray}
cf. \cite{ps}. This equation holds for all $\bm x$ on $S_{\alpha}$ with $\alpha=1, 2$. Together with the conditions of zero forces and torques it determines $\bm V_{\alpha}$, $\bm \Omega_{\alpha}$ and the surface traction
uniquely \cite{kim}. We use the assumption $h\gg z_{\alpha}$, meaning that the wall at $z=h$ is much further from the spheres than the one at $z=0$. We can therefore approximately assume
\begin{eqnarray}&&\!\!\!\!\!\!\!\!\!\!\!\!\!\!\!\!
S_{il}(\bm x, \bm x')\approx G_{il}(\bm x, \bm x'), \label{condi}
\end{eqnarray}
where $G_{il}(\bm x, \bm x')$ is the Stokeslet near a plane wall defined by
\begin{eqnarray}&&
-\nabla p'+\eta \nabla^2 \bm u'+\bm g\delta(\bm x-\bm x')=0,\ \ \nabla\cdot\bm u'=0, \ \ \bm u'(z=0)=\bm u'(x\to\infty)=0,\ \
\bm u'(\bm x)=\frac{G_{il}(\bm x, \bm x')g_l}{8\pi \eta}.\label{stokes}
\end{eqnarray}
The requirement that Eq.~(\ref{condi}) holds when both $\bm x$ and $\bm x'$ belong to the spheres quantifies the assumption that one of the walls is much further than the other. In practice the difference between distances to the upper and lower walls does not have to be too large for the equation to hold. With $z_{\alpha}\ll h$ and this assumption, Eq.~(\ref{intr1}) becomes
\begin{eqnarray}&&\!\!\!\!\!\!\!\!\!\!\!\!\!\!\!\!
(V_{\alpha})_i+(\bm \Omega_{\alpha}\times (\bm x-\bm x_{\alpha}))_i\!=\!\dot{\gamma}\delta_{ix}z
-\!\sum_{\alpha'}\int_{S_{\alpha'}}\!\!\!\frac{G_{il}(\bm x, \bm x')t_{l}(\bm x')dS'}{8\pi \eta}. \label{intr}
\end{eqnarray}
We next introduce the decomposition of $G_{il}(\bm x, \bm x')$ into the Stokeslet in an infinite space and the correction due to the wall ${\tilde G}_{il}$,
\begin{eqnarray}&&\!\!\!\!\!\!\!\!\!\!\!\!\!\!
G_{il}(\bm x, \bm x')=Y_{il}(\bm r)+{\tilde G}_{il}(\bm x, \bm x'), \label{do}
\end{eqnarray}
with $\bm r=\bm x-\bm x'$, as above. The contribution ${\tilde G}_{il}$ is induced by the images located at the reflection $(\bm x')^*=(x', y', -z')$ of the source position $\bm x'=(x', y', z')$ with respect to the plane $z=0$. It was found in \cite{Blake} that the image singularities are a point force of the same magnitude as the source, but with an opposite sign, a stokes-doublet and a source-doublet, see definitions in the paper. We can write the formula in Ref.~\cite{Blake} as follows
\begin{eqnarray}&&\!\!\!\!
{\tilde G}_{il}(\bm x, \bm x')=-Y_{il}(\bm R)+2z'G^{(1)}_{il}(\bm R)+2z'^2 G^{(2)_{il}}(\bm R),\ \ G^{(1)}_{il}\!=\!\left(2\delta_{3l}\!-\!1\right)\partial_l Y_{i3},\ \
G^{(2)}_{il}\!=\!\frac{\left(1\!-\!2\delta_{3l}\right)\left(R^2\delta_{il}\!-\!3R_iR_l\right)}{R^5},\label{gr}
\end{eqnarray}
where $\bm R=\bm x-(\bm x')^*$ is the distance from the images and there is no summation over repeated indices. The symmetries of the Green's functions $G_{il}(\bm x, \bm x')=G_{li}(\bm x', \bm x)$ and $Y_{il}(\bm x, \bm x')=Y_{li}(\bm x', \bm x)$ imply the symmetry ${\tilde G}_{il}(\bm x, \bm x')={\tilde G}_{il}(\bm x, \bm x')$, which can be confirmed directly.

We compare Eq.~(\ref{intr}) with the similar equation for two spheres driven by the unbounded shear flow that was considered above. The equation can be obtained by dropping ${\tilde G}_{il}$ above, see Eq.~(\ref{inter}), which yields
\begin{eqnarray}&&\!\!\!\!\!\!\!\!\!\!\!\!\!\!\!\!
(V^0_{\alpha})_i\!+\!(\bm \Omega^0_{\alpha}\!\times\! (\bm x\!-\!\bm x_{\alpha}))_i\!=\!\dot{\gamma}\delta_{ix}z
-\!\sum_{\alpha'}\int_{S_{\alpha'}}\!\!\!\frac{Y_{il}(\bm x\!-\!\bm x')t^0_{l}(\bm x')dS'}{8\pi \eta}.\label{intb}
\end{eqnarray}
Subtracting Eq.~(\ref{intb}) from Eq.~(\ref{intr}) we find
\begin{eqnarray}&&\!\!\!\!\!\!\!
(\delta V_{\alpha})_i\!+\!(\delta\bm \Omega_{\alpha}\!\times\! (\bm x\!-\!\bm x_{\alpha}))_i\!+\!\sum_{\alpha'}\int_{S_{\alpha'}}\!\!\!\frac{{\tilde G}_{il}(\bm x, \bm x')t^0_{l}(\bm x')dS'}{8\pi \eta}\!=\!-\!\sum_{\alpha'}\int_{S_{\alpha'}}\!\!\!\frac{G_{il}(\bm x, \bm x')\delta t_{l}(\bm x')dS'}{8\pi \eta}.\label{ineg}
\end{eqnarray}
Provided that $\bm t^0(\bm x)$ is known, this is an integral equation on $\delta\bm t(\bm x)$ which also obeys the conditions of zero forces and torques given by Eqs.~(\ref{inertless}) with $\delta \bm t$ replacing $\bm t$. So far we have not made any approximations besides that the spheres are much closer to one of the two walls of the channel. \\ \\

\subsection{Asymptotic solution for a distant wall}

We consider the solution of Eq.~(\ref{ineg}) in the limit of a distant wall.
The last (source) term in the left hand side of this equation, in contrast with the rest of the terms, does not involve properties of $\delta\bm u$.
If it is dropped, then we find Eq.~(\ref{intr}) with $\dot{\gamma}=0$, that is the equation for two inertialess spheres moving near the wall in the fluid at rest, which unique solution is trivial -- zero translational and angular velocities.

In fact, Eq.~(\ref{ineg}) coincides with the equation for the velocities of an inertialess swimmer, composed of two spheres, that swims near a plane wall at $z=0$. In this case, the propulsion is powered by the swimming stroke prescribed by the velocity distribution at the spheres' surface as given by that last term.

When both spheres are separated from the wall by a distance much larger than $a$ the asymptotic series solution can be obtained via the Taylor expansion of ${\tilde G}_{il}(\bm x, \bm x')$ near the centers of the spheres, cf.
Sec.~\ref{direct}. Indeed, both arguments, $\bm x$ and $\bm x'$, of ${\tilde G}_{il}(\bm x, \bm x')$ are confined in Eq.~(\ref{ineg}) to one of the spheres (possibly different ones).
In this range ${\tilde G}_{il}(\bm x, \bm x')$ is a slowly varying function of its arguments because $z_{\alpha}\gg a$ and the image of $\bm x'$ under this condition is separated from each sphere by a distance much larger than the radius, cf. with the Lorentz solution \cite{hb} and also Appendix of Ref.~\cite{fl2018}.
This observation does not depend on the separation between the spheres that can be nonetheless arbitrary.
Thus we write Eq.~(\ref{ineg}) as,
\begin{eqnarray}&&\!\!\!\!\!\!\!
(\delta V_{\alpha})_i\!+\!(\delta\bm \Omega_{\alpha}\!\times\! (\bm x\!-\!\bm x_{\alpha}))_i\!+\!\sum_{\alpha'}\int_{S_{\alpha'}}\!\!\!\frac{{\tilde G}_{il}(\bm x_{\alpha}, \bm x')t^0_{l}(\bm x')dS'}{8\pi \eta}
+(\bm x\!-\!\bm x_{\alpha})_k\frac{\partial}{\partial x_k}\sum_{\alpha'}\int_{S_{\alpha'}}\!\!\!\frac{{\tilde G}_{il}(\bm x, \bm x')t^0_{l}(\bm x')dS'}{8\pi \eta}|_{\bm x=\bm x_{\alpha}}+\ldots
\nonumber\\&&\!\!\!\!\!\!\!
=\!-\!\sum_{\alpha'}\int_{S_{\alpha'}}\!\!\!\frac{G_{il}(\bm x, \bm x')\delta t_{l}(\bm x')dS'}{8\pi \eta}, \label{fdk}
\end{eqnarray}
where dots stand for higher order terms in the Taylor expansion. The asymptotic solution can be obtained by requiring that the equation holds at every order in $\max [a/z_1, a/z_2]$ (the case of disparate $z_{\alpha}$ seems to be of little interest so $z_1\sim z_2$ can be assumed below though this is not necessary for the analysis). The zero-order term determines the particle velocities,
\begin{eqnarray}&&\!\!\!\!\!\!\!
(\delta V_{\alpha})_i=\!-\!\sum_{\alpha'}\int_{S_{\alpha'}}\!\!\!\frac{{\tilde G}_{il}(\bm x_{\alpha}, \bm x')t^0_{l}(\bm x')dS'}{8\pi \eta}, \label{ordsa}
\end{eqnarray}
where $\delta t_{l}$ is zero at this order. This formula can be simplified by noting that ${\tilde G}_{il}(\bm x_{\alpha}, \bm x')$ is a smooth function of $\bm x'$ on each of the spheres for the same reasons as before due to the symmetry ${\tilde G}_{il}(\bm x, \bm x')={\tilde G}_{li}(\bm x', \bm x)$. The zero-order term in the expansion vanishes by the condition of zero force. We thus find that
\begin{eqnarray}&&\!\!\!\!\!\!\!\!\!\!\!\!\!\!\!\!
\delta V_{\alpha i}=-\frac{S_{lm}}{8\pi \eta}\sum_{\alpha'}\frac{\partial {\tilde G}_{il}(\bm x_{\alpha}, \bm x_{\alpha'})}{\partial (x_{\alpha'})_m}+o\left(\frac{a}{z_{\alpha}}\right), \label{response}
\end{eqnarray}
which is a more rigorous derivation of the result that might also be obtained using reflections as described in the beginning of the Section. The use of an integral representation allows us to precisely formulate the validity conditions and to provide a transparent structure of the asymptotic series. It is important for the further analysis that the derivation does not assume any \emph{a priori} relation between $\bm V^0_{\alpha}$ and $\delta \bm V_{\alpha}$. Actually, the absolute value of the velocities $\delta \bm V_{\alpha}$ would be smaller than $\bm V^0_{\alpha}$, however, this need not to be true for the relative velocities which are of main interest here.

\subsection{Evolution equation of inter-particle distance} \label{dista}

The velocity of the relative motion of the spheres is described by $\bm V=\bm V_2-\bm V_1$ that, at the leading order, obeys
\begin{eqnarray}&&\!\!\!\!\!\!\!\!\!\!\!\!\!\!\!
V_i\!=\!V^0_i\!+\!\frac{S_{lm}}{8\pi \eta}\sum_{\alpha'}\left(\frac{\partial {\tilde G}_{il}(\bm x_1, \bm x_{\alpha'})}{\partial (x_{\alpha'})_m}\!-\!\frac{\partial {\tilde G}_{il}(\bm x_2, \bm x_{\alpha'})}{\partial (x_{\alpha'})_m}\right). \label{vls}
\end{eqnarray}
The relative velocity in an unbounded shear flow $\bm V^0$ can be written as \cite{ujhd}
\begin{eqnarray}&&\!\!\!\!\!\!\!\!\!\!\!\!\!\!\!\!
V^0_i(\bm r)\!=\!\dot \gamma z\delta_{i1}\!-\!\frac{\dot \gamma B  z\delta_{i1}}{2}\!-\!\frac{\dot \gamma B x\delta_{i3}}{2}\!-\!\frac{\dot \gamma(A-B)xz r_i}{r^2}, \label{unpert}
\end{eqnarray}
where $\bm r=\bm x_2-\bm x_1$. The first term in the RHS is the driving shear flow. The remaining terms, due to hydrodynamic interactions, are described by the functions $A$ and $B$, which depend on $|\bm r|=r$ only. These functions are considered, similarly to $K$, $L$ and $M$ above, as given \cite{ujhd}.
We observe that $\bm V^0$ is determined uniquely by the distance between the particles and is independent of the particles' center of mass. Thus the evolution of $\bm r(t)$ without the wall is autonomous, i.e., the time derivative of $\bm r(t)$ is determined uniquely by the instantaneous value of $\bm r(t)$. We demonstrate that the
evolution of $\bm r(t)$, described by Eq.~(\ref{vls}), remains autonomous. This means that we can neglect in $\bm V$, which is a function of $\bm x_i$, the dependence on the center-of-mass coordinate $(\bm x_1+\bm x_2)/2$. Since the horizontal coordinates of the center of mass are irrelevant by translational invariance in the horizontal directions, we need to consider only the dependence on $z_0=(z_1+z_2)/2$. This coordinate would not change at all without the hydrodynamic interactions and the particles would move in straight lines parallel to the wall. The interactions cause temporal variations of $z_0$; however these occur only over the scale of these interactions which is the radius $a$. Moreover, this change is small by the assumption that $\max [a/z_1, a/z_2]\ll 1$. This allows to consider $z_0=(z_1+z_2)/2$ as constant during the whole time of the interactions giving $\bm V=\bm V(\bm r(t), z_0(t))\approx \bm V(\bm r(t), z_0(t_0))$ where $t_0$ is arbitrary. For $\delta V_i\equiv V_i\!-\!V^0_i$ we have
\begin{eqnarray}&&
\delta V_i\!=\!\frac{S_{lm}}{8\pi \eta}\sum_{\alpha'}\left(\frac{\partial {\tilde G}_{li}(\bm x_{\alpha'}, \bm x_1)}{\partial (x_{\alpha'})_m}\!-\!\frac{\partial {\tilde G}_{li}(\bm x_{\alpha'}, \bm x_2)}{\partial (x_{\alpha'})_m}\right)
=\!\frac{S_{lm}}{8\pi \eta}\sum_{\alpha'}
\frac{\partial}{\partial (x_{\alpha'})_m}
\left(Y_{li}(\bm x_{\alpha'}-\bm x_2^*)-Y_{li}(\bm x_{\alpha'}-\bm x_1^*)
+2z_1G^1_{li}(\bm x_{\alpha'}-\bm x_1^*)\right.\nonumber\\&&\left.
+2z_1^2G^2_{li}(\bm x_{\alpha'}-\bm x_1^*)-2z_2G^1_{li}(\bm x_{\alpha'}-\bm x_2^*)
-2z_2^2G^2_{li}(\bm x_{\alpha'}-\bm x_2^*)
\right),\label{dfr}
\end{eqnarray}
where we used Eq.~(\ref{gr}). The derivatives in the above equation can be written via the tensors
\begin{eqnarray}&&\!\!\!\!\!
T^{(1)}_{lim}(\bm r)\!\equiv\! -\frac{\partial Y_{li}(\bm r)}{\partial r_m}\!=\!\frac{r^2\left(r_m\delta_{il}\!-\!r_i\delta_{lm}\!-\!r_l\delta_{im}\right)
\!+\!
3r_ir_lr_m}{r^5};\ \
T^{(2)}_{lim}(\bm r)\!\equiv\!\frac{\partial  G^1_{li}}{\partial r_m}
=\left(1\!-\!2\delta_{3i}\right)\left(\frac{\delta_{im}\delta_{3l}\!-\!\delta_{i3}\delta_{lm}\!-\!\delta_{il}\delta_{3m}}{r^3}
\right.\nonumber\\&&\!\!\!\!\!\left.-\frac{3r_i(r_m\delta_{3l}\!-\!r_3\delta_{lm}\!-\!r_l\delta_{3m})}{r^5}
+\frac{3(\delta_{i3}r_lr_m\!+\!\delta_{il}r_3r_m\!+\!\delta_{im}r_3r_l)}{r^5}\!-\!\frac{15r_ir_3r_lr_m}{r^7}
\right);\nonumber\\&&\!\!\!\!\!
T^{(3)}_{lim}(\bm r)\!\equiv\!\frac{\partial  G^2_{li}}{\partial r_m}
\!=\!-3\!\left(1\!-\!2\delta_{3i}\right)\left(\frac{r_m\delta_{il}\!+\delta_{im}r_l\!+\!\delta_{lm}r_i}{r^5}\!
-\!\frac{5r_mr_ir_l}{r^7}
\right),
\label{tensors}
\end{eqnarray}
where we used Eqs.~(\ref{do}) and (\ref{gr}). We also observe that,
\begin{eqnarray}&&\!\!\!\!\!\!\!\!
\bm x_{\alpha'}-\bm x_{\alpha}^*=(x_{\alpha'}-x_{\alpha}, y_{\alpha'}-y_{\alpha}, z_{\alpha'}+z_{\alpha}).
\end{eqnarray}
Thus we find,
\begin{eqnarray}&&\!\!\!\!\!\!\!\!\!\!\!\!\!\!\!\!
\delta V_i
=\!\frac{S_{lm}}{8\pi \eta}
\left[K_{lim}^{11}-K_{lim}^{12}+K_{lim}^{21}-K_{lim}^{22}
\right];\ \ K_{lim}^{kn}=
T^{(1)}_{lim}(\bm x_k-\bm x_n^*)+2z_jT^{(2)}_{lim}(\bm x_k-\bm x_n^*)+2z_j^2T^{(3)}_{lim}(\bm x_k-\bm x_n^*).
\label{fk}
\end{eqnarray}
We write above $z_2=z_0+r_3/2$ and $z_1=z_0-r_3/2$ where $r_3=z_2-z_1$ is the vertical component of the distance $\bm r$.
We use $\bm x_{\alpha}-\bm x_{\alpha}^*=2z_{\alpha}$ so that,
\begin{eqnarray}&&\!\!\!\!\!\!\!\!\!\!\!\!\!\!\!\!
T^{(k)}_{lim}(\bm x_1-\bm x_1^*)=T^{(k)}_{lim}(0, 0, 2z_0-r_3);\ \
T^{(k)}_{lim}(\bm x_2-\bm x_2^*)=T^{(k)}_{lim}(0, 0, 2z_0+r_3). \label{tesnor}
\end{eqnarray}
Similarly, using $\bm x_1-\bm x_2^*=(-r_1, -r_2, z_1+z_2)=(-r_1, -r_2, 2z_0)$ and $\bm x_2-\bm x_1^*=(r_1, r_2, 2z_0)$ we find,
\begin{eqnarray}&&\!\!\!\!\!\!\!\!\!\!\!\!\!\!\!\!
T^{(k)}_{lim}(\bm x_1-\bm x_2^*)=T^{(k)}_{lim}(-r_1, -r_2, 2z_0);\ \
T^{(k)}_{lim}(\bm x_2-\bm x_1^*)=T^{(k)}_{lim}(r_1, r_2, 2z_0).\label{tesnor1}
\end{eqnarray}
The last equations provide the velocity in Eq.~(\ref{fk}) in terms of $\bm r$ and $z_0$.
We consider $z_0$ as a constant given by the initial configuration, see the discussion
after Eq.~(\ref{unpert}). The remaining terms in Eq.~(\ref{dfr}) depend only on $\bm r$, providing an autonomous equation for $\bm r$.

The detailed form of the evolution equation for $\bm r$ in Cartesian coordinates is given by Eqs.~(\ref{delV})-(\ref{delV1}) in Appendix \ref{cartesian}. The more compact form is found by employing the cylindrical coordinate system with $x = \rho \cos\phi,\ y = \rho \sin\phi,\ z=z$. We find using the identities $x \dot x + y \dot y = \rho \dot\rho$ and $-y \dot x + x \dot y = \rho^2 \dot\phi$ and the definitions $s^2\equiv \rho^2+4z_0^2$ and $\sigma\equiv r^2-s^2$ that (here and below we set $\dot{\gamma}=1$ by passing in the equation of
motion for $\bm r$ to dimensionless time $\dot{\gamma} t$),
\begin{eqnarray}&&\!\!\!\!\!\!\!\!\!
\dot\rho = z c_{\phi}
\left[
1-\frac{B}{2}-\frac{\rho^2(A-B)}{r^2}
+\frac{5\rho^2 P}{3r^4s^5}
+\frac{10z_0 R}{r^4\sigma^2}
+\frac{5\rho^2\sigma}{2r^4s^7}
(P+2(\rho^4-s^2\rho^2+4z^2z_0^2)M)
\right].
\label{rho}
\end{eqnarray}
Here, we have introduced $c_{\phi} = \cos\phi$, $P = r^2s^2(L-M)+3(\rho^4+4z^2z_0^2)M$ and
$R = r^4(1+K+L)+2z^2\rho^2M$, see definitions in Eqs.~(\ref{dipole}) and (\ref{unpert}). The dynamics of $c_{\phi}$ is
\begin{eqnarray}&&\!\!\!\!\!\!\!\!\!\!\!\!\!\!\!\!\!\!\!\!
\dot c_{\phi} = \frac{z}{\rho}(c_{\phi}^2-1)\left[
\frac{B}{2}-1+
\frac{5\rho^2\sigma L}{2r^2s^5}
-\frac{10z_0}{r^2\sigma^2}(r^2(1+K)+z^2L)
\right].
\label{phi1}
\end{eqnarray}
Finally, the dynamics of $z$ reads
\begin{eqnarray}&&
\dot z = \rho c_{\phi}
\left[
-\frac{B}{2}-\frac{z^2(A-B)}{r^2}
+\frac{5(\rho^2-16z_0^2)\sigma R}{2r^4s^7}
+\frac{5z^2 P}{3r^4s^5}
+\frac{10z^2 z_0}{r^4\sigma^2}(r^2 L+(2z^2-\rho^2)M)
\right].
\label{zz}
\end{eqnarray}
Further noting that $\rho^4-s^2\rho^2+4z^2z_0^2 =
\rho^2(\rho^2-s^2)+4z^2z_0^2 =
-4z_0^2\rho^2+4z^2z_0^2 =
4z_0^2(z^2-\rho^2)$ the evolution equation for $\rho$ can be rewritten as
\begin{eqnarray}&&\!\!\!\!\!\!\!\!\!\!
\dot\rho = z c_{\phi}
\left[
1-\frac{B}{2}-\frac{\rho^2(A-B)}{r^2}
+\frac{5\rho^2 P}{3r^4s^5}
+\frac{10z_0 R}{r^4\sigma^2}
+\frac{5\rho^2\sigma P}{2r^4s^7}
+\frac{20\rho^2 z_0^2\sigma}{r^4s^7}
(z^2-\rho^2)M
\right].
\label{rho2}
\end{eqnarray}
It can be readily seen using $|s|\sim z_0$ and $|\sigma| \sim z_0^2$, that at fixed $\bm r$ we have $\delta V_i \sim z_0^{-3}$ upon varying $z_0$. Similarly if we fix $z_0$ then $\delta V_i \sim r^{-2}$ upon varying $r$. The inverse cubic dependence on $z_0$ is non-trivial. Derivatives of ${\tilde G}_{il}(\bm r)$ contain terms of order $r^{-2}$ which would give $z_0^{-2}$ behavior in Eq.~(\ref{response}), cf. the dependence of $T^k_{lim}$ on $r$ in Eq.~(\ref{tensors}). Following rules for tensorial transformations upon the sign reversal of the argument, see, {\it e. g.}, Eqs.~(\ref{tesnor})-(\ref{tesnor1}), the leading order $z_0^{-2}$ terms cancel.

We remark that finding the next order correction to $\bm V$ in the inverse distance to the wall would involve the quadratic surface moments originating from $\bm t^0$. These were not considered previously and would be quite demanding to compute, see Eq.~(\ref{ordsa}). It would also require considering the contributions in the second line of Eq.~(\ref{fdk}). The corresponding exceedingly complex calculations are beyond the scope of the present paper. We take here the practical approach of trying to push our leading order calculation to smaller $z_0$ and compare the analytical prediction with the results of the direct numerical simulations.

The equations of motion have symmetries that can be described as the properties of the velocity components,
\begin{eqnarray}&&\!\!\!\!\!\!\!\!\!\!
V_{\rho}(\rho,-\phi,z)\!=\!V_{\rho}(\rho,\phi,z),\
V_{\rho}(\rho,\phi,-z)\!=\!-V_{\rho}(\rho,\phi,z),\ \
V_{\phi}(\rho,-\phi,z)\!=\!V_{\phi}(\rho,\phi,z),\
V_{\phi}(\rho,\phi,-z)\!=\!-V_{\phi}(\rho,\phi,z),
\nonumber\\
&&\!\!\!\!\!\!\!\!\!\!
V_{z}(\rho,-\phi,z)\!=\!V_{z}(\rho,\phi,z),\
V_{z}(\rho,\phi,-z)\!=\!V_{z}(\rho,\phi,z).
\label{symmetriesCyl}
\end{eqnarray}
These properties allow to confine the study of the trajectories $\bm r(t)$ to $z \ge 0,\ 0\le\phi\le\pi/2$, besides the constraint $r\ge 2$.

The main result of this Section is the evolution equation for the distance between two spheres freely suspended in a shear flow near the wall,
\begin{eqnarray}&&\!\!\!\!\!\!\!\!\!\!\!\!\!\!\!\!
\dot{\bm r}=\bm V(\bm r)=\bm V^0(\bm r)+\delta \bm V(\bm r).
\end{eqnarray}
Here $\bm V^0(\bm r)$, given by Eq.~(\ref{unpert}), describes hydrodynamic interactions due to shear in unbounded flow and $\delta \bm V$ describes the effects of the wall, given at the leading order
by Eqs.~(\ref{delV})-(\ref{delV1}). Despite that the wall is assumed to be distant, its effect
is not small even for large channels.

\section{Singular effect of the wall at far distances}\label{singular}

In this Section we demonstrate that the wall is a singular perturbation of the relative motion between the two spheres. Regardless of how large $z_0$ is, its influence cannot be entirely neglected.
For any fixed $\bm r$ we have $\bm V(\bm r)=\bm V^0(\bm r)$ for $z_0\to\infty$. However for any fixed $z_0\gg 1$ there are large $\bm r$ for which some velocity components satisfy $|\delta V_i|\gg V^0_i$. There is a competition between the different parameters: the hydrodynamic interactions are small by $a/r$ whereas the interaction with the wall is small by $a/z_0$. As a result at $r$ given by a power of $z_0$, whose exponent is determined by the details of the power laws of the particle-particle and particle-wall interactions, the interactions with the wall may dominate the evolution of $\bm r$. The resulting topology of the trajectories of the relative motion is hence different, as we will describe in the next Sections. First, we illustrate the differences numerically.

\begin{figure*}
\begin{center}
\begin{tabular}{cc}
\includegraphics[width=0.45\textwidth]{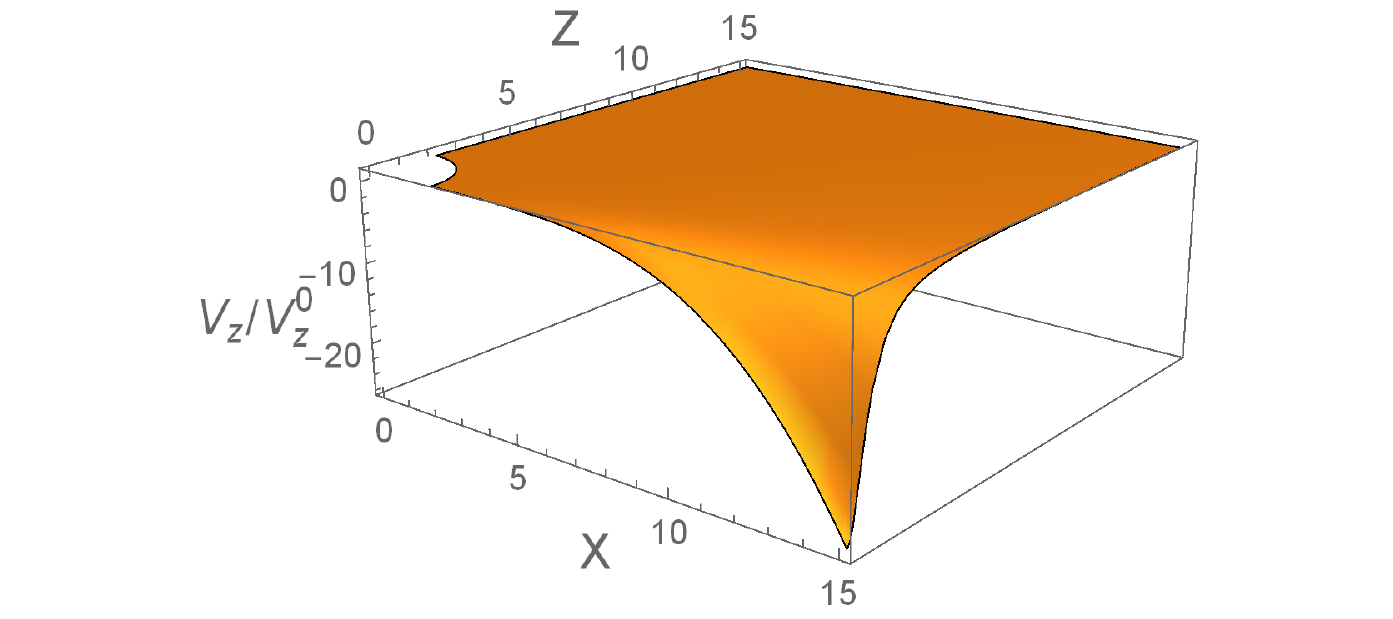} &
\includegraphics[width=0.4\textwidth]{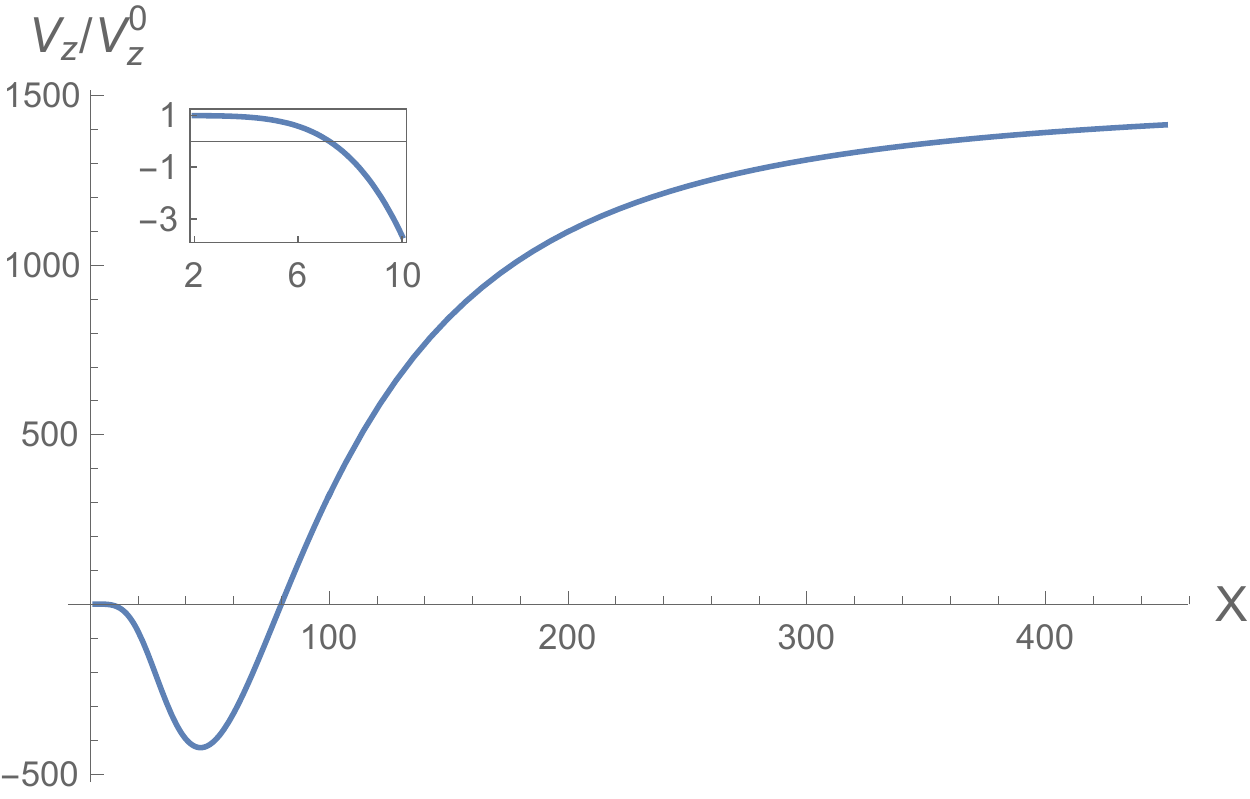}
\\
(a) & (b)
\end{tabular}
\end{center}
\caption{The ratio $V_z/V_z^0$ for $z_0=20$  (a) in the $xz$-plane and (b) along the $x$-axis. The presence of the wall increases
the velocity at $x\gg z_0$ by a constant large factor of order $z_0^2$. The two critical points are the neutral equilibrium point $r_c=4z_0=80$
and the saddle point $r_s=(32 z_0^3/15)^{1/5}$ (see the inset in (b)).}
\label{Ratio1}
\end{figure*}

Trajectories that pass through points with $y=0$ belong to  the $xz$-plane by symmetry, as $V_y(y=0)=0$. We consider the remaining components $V_x$ and $V_z$ as functions of $x$ and $z$ in the $xz$-plane. We can restrict the analysis to positive $x$ and $z$ due to the symmetries described in the previous Section. For the streamwise component of the velocity the wall is a regular perturbation: the ratio $V_x/V_x^0$ is everywhere close to 1.  Thus for $z_0=20$ the maximal deviation of $V_x/V_x^0$ from 1 is seen numerically to be less than 1\%. Consider for instance the ratio at $x=y=0$ where the only nonzero components are $V_x^0(0, 0, z)=(1-B(z)/2)z$ and $\delta V_x(0, 0, z) = 10 z z_0(1+K(z)+L(z))/(z^2-4z_0^2)^2$, see
Eqs.~(\ref{unpert}) and (\ref{delV}). When $z \sim 2$ all $B(z),K(z),L(z)$ are finite and less than unity \cite{ujhd}. By taking the ratio we find that $\delta V_x/V_x^0 \sim  10/z_0^3\ll 1$ at $z_0\gg 1$.
At large distances $V_x^0$ is dominated by the driving shear flow, see Eq.~(\ref{unpert}), and it is much larger than $\delta V_x$ because the symmetry imposes proportionality of $\delta V_x$ and $z$, see Eq.~(\ref{delV}).
Similarly, in other cases, $V_x/V_x^0\approx 1$. Thus for practical purposes we can set
\begin{eqnarray}&&\!\!\!\!\!\!\!\!\!\!\!\!\!\!\!\!
V_x(\bm r)\approx V^0_x(\bm r)\!=\!\left(1\!-\!\frac{B(r)}{2}\right) z\!-\!\frac{(A(r)-B(r))x^2z}{r^2}, \label{practical}
\end{eqnarray}
which at large distances reduces to the carrying shear flow difference given by $z$.
The situation is quite different for $V_z/V_z^0$ in Fig. \ref{Ratio1}.
We see that when the spheres are close, the difference is negligible and $V_z\approx V_z^0$. However the situation is quite different at
large separations. From Eq.~(\ref{unpert}) we obtain that for a wall at infinite distance,
\begin{eqnarray}&&\!\!\!\!\!\!\!\!\!\!\!\!\!\!\!\!
V^0_z(\bm r)\!=\!-\frac{ B x }{2}\!-\!\frac{(A-B)xz^2}{r^2},\label{sdh}
\end{eqnarray}
which can be further simplified at large $r$ using
\begin{eqnarray}&&\!\!\!\!\!\!\!\!\!\!\!\!\!\!\!\!
A(r)=\frac{5}{r^3}+O\left(\frac{1}{r^5}\right),\ \ B(r)=\frac{16}{3r^5}+o\left(\frac{1}{r^6}\right), \label{largr}
\end{eqnarray}
see \cite{ujhd}. We thus find that
\begin{eqnarray}&&\!\!\!\!\!\!\!\!\!\!\!\!\!\!\!\!
V^0_z(\bm r)\!\approx \!-\frac{8x }{3r^5}\!-\!\frac{5 xz^2}{r^5},
\label{bgz}
\end{eqnarray}
which is negative at $x>0$. We observe from Eqs.~(\ref{largr}) that the order of corrections is quite high so this formula might hold already at $r\simeq 3-4$.
For $\delta V_z$ we have from the equations in Appendix \ref{cartesian} that
\begin{eqnarray}&&\!\!\!\!\!\!\!\!\!\!\!\!\!\!\!\!
\delta V_z(x, 0, 0)\! =\!\frac{10x z_0^2(16z_0^2\!-\!x^2)}{(x^2+4z_0^2)^{7/2}}(1+K(x)+L(x)). \label{delata}
\end{eqnarray}
This can be simplified at large $x$ following Ref.~\cite{ujhd},
\begin{eqnarray}&&\!\!\!\!\!\!\!\!\!\!\!\!\!\!\!\!
K(r)\!\approx \!-\frac{2}{r^5},\ \ L(r)\!\approx \!-\frac{5}{2r^3},\ \ M(r)\!\approx \!\frac{25}{2r^3},\ \ r\gg 1, \label{dsf}
\end{eqnarray}
by neglecting $K$ and $L$ compared to unity in Eq.~(\ref{delata}). These functions decay fast with $r$ implying that
\begin{eqnarray}&&\!\!\!\!\!\!\!\!\!\!\!\!\!\!\!\!
\delta V_z(x, 0, 0)\!\approx \!\frac{10x z_0^2(16z_0^2\!-\!x^2)}{(x^2+4z_0^2)^{7/2}}, \label{velsc}
\end{eqnarray}
must hold already at $x=3$ where the spheres are rather close. We also find for the ratio of velocities
\begin{eqnarray}&&\!\!\!\!\!\!\!\!\!\!\!\!\!\!\!\!
\frac{\delta V_z(x, 0, 0)}{V^0_z(x, 0, 0)}\! =\!\frac{15x^5z_0^2(x^2\!-\!16z_0^2)}{4(x^2+4z_0^2)^{7/2}};\ \ \ x\gg 1.\label{fok}
\end{eqnarray}
The corrections are of order higher than $1/x$ so in practice this formula works at rather small $x$.

The ratio on the LHS of Eq.~(\ref{fok}) equals $-1$ at $x$ obeying the condition,
\begin{eqnarray}&&\!\!\!\!\!\!\!\!\!\!\!\!\!\!\!\!
15x^5z_0^2(x^2\!-\!16z_0^2)+4(x^2+4z_0^2)^{7/2}=0.
\label{critical_pts_eq}
\end{eqnarray}
This equation has two solutions. The first one is obtained when
$x \ll 2z_0$ and hence $15x^5 = 32 z_0^3$ leading to $r_s = (32 z_0^3/15)^{1/5}$. This expression for the critical point, obtained from the $x\gg 1$ approximation given by Eq.~(\ref{fok}), is indistinguishable from the numerical solution of $V_z(x, 0, 0)=0$ with the full velocity given by Eqs.~(\ref{delV})-(\ref{delV1}), at least down to $z_0=5$ which is the smallest $z_0$ considered in this work (we have $r_s\approx 3$ at $z_0=5$). This is reasonable in view of the remarks after Eqs.~(\ref{bgz}),  (\ref{velsc}) and (\ref{fok}). To find the other solution we notice that $x^2=16z_0^2+\delta$ with $\delta \ll 16z_0^2$ solves the equation giving $r_c \approx 4z_0$. Both $r_s$ and $r_c$ are much larger than unity at $z_0\gg 1$ confirming the consistency of the approach and can be used for $z_0\geq 5$.

The obtained points obey $V_z(r_c, 0, 0)=V_z(r_s, 0, 0)=0$. Moreover $V_x(x, 0, 0)=V_y(x, 0, 0)=0$ since both $V_x(x, y, z)$ and $V_y(x, y, z)$ are odd functions of $z$, see Eq.~(\ref{symmetries}). Thus, the points on the $x$-axis with $x=r_s$ and $x=r_c$ are the critical points with $\bm V=0$. We demonstrate below that these are a saddle point and a stationary point, respectively.

Finally we would like to emphasise the singular nature of the perturbation due to the long-range interaction at finite $z_0$. For motions in the plane $y=0$, at large but finite $z_0$, there are locations $\bm r$ for which the wall-normal component of the velocity $V_z$ is much larger than the BG velocity, see the $z_0=20$ case in Fig. \ref{Ratio1}. In fact, for $x\gg z_0$, the ratio $\delta V_z(x, 0, 0)/V^0_z(x, 0, 0)$ becomes an $x-$independent constant of order $z_0^2\gg 1$. The wall contribution to the velocity is opposite in sign to the BG velocity. If we consider two particles on the same streamline of the unperturbed flow with $y=z=0$, then the only non-vanishing, $z$-component of
the velocity $V_z(x, 0, 0)$ is,
\begin{eqnarray}&&\!\!\!\!\!\!\!\!\!\!\!\!\!\!\!\!
V_z\!=\!\frac{10x z_0^2(16z_0^2\!-\!x^2)}{(x^2\!+\!4z_0^2)^{7/2}}(1\!+\!K(x)\!+\!L(x))\!-\!\frac{xB(x)}{2}. \label{frac}
\end{eqnarray}
In the BG limit of $z_0\to \infty$, taken at fixed $x$, the first term drops, reducing the velocity to $V_z^0(x, 0, 0)=xB(x)/2$, see Eq.~(\ref{sdh}), and at large distances  $xB(x)/2\approx 8/(3x^{4})$, see Eq.~(\ref{largr}). In contrast, at any finite $z_0$, for $x\gg z_0$, the range not considered in the BG approximation, the contribution due to the wall, described by the first term in Eq.~(\ref{frac}) behaves as $z_0^2 x^{-4}$. We find, using that the functions $K(x)$ and $L(x)$  vanish at large distances by Eq.~(\ref{dsf}),
\begin{eqnarray}&&\!\!\!\!\!\!\!\!\!\!\!\!
\lim_{x\to\infty}\frac{\delta V_z(x, 0, 0)}{V_z^0(x, 0, 0)}\!=\!\lim_{x\to\infty}\frac{20 z_0^2(x^2\!-\!16z_0^2)}{B(x)(x^2\!+\!4z_0^2)^{7/2}}
=\frac{15 z_0^2}{4}. \label{limit}
\end{eqnarray}
Thus the interaction between particles flowing along the same streamline is dominated by the wall term at $x\gtrsim z_0$. This sets in non-uniformly. We see
from Fig. \ref{Ratio1}(b)
that for $z_0=20$ the absolute value $|V_z(x, 0, 0)/V_z^0(x, 0, 0)|$ grows fast with $x$. It crosses zero (which corresponds to $|\delta V_z(x, 0, 0)/V^0_z(x, 0, 0)|=1$) at the critical saddle point $(32 z_0^3/15)^{1/5}\approx 7$, a value smaller than the half of $z_0$. One might have expected that $r\lesssim z_0$ guarantees at least a qualitative validity of the BG theory, however it does not. The ratio $|\delta V_z(x, 0, 0)/V^0_z(x, 0, 0)|$ rapidly grows with $x$, becoming of order one hundred already at $x \approx 30$. However, after reaching the maximum, it decreases to the value 1 at the critical point at $x=4z_0$. Only at $x\gg 4z_0$ the asymptotic law $|V_z(x, 0, 0)/V_z^0(x, 0, 0)| \sim z_0^2$ starts to apply. We find numerically that the curve $|V_z(x, 0, 0)/V_z^0(x, 0, 0)|$ starts flattening at $x\sim 200$ when its value is about one thousand. The approach to the limiting value of $1500$, imposed by Eq.~(\ref{limit}), is quite slow: e.g.\ at $x\simeq 450$ the ratio is about $1400$. We conclude that, at the considered value of $z_0$, the wall dominates the interactions at all $x\gtrsim z_0/2$, excluding a small neighborhood of the neutral equilibrium critical point $r_s$.

The strong changes of $\bm V$ induced by the presence of a wall described in this Section imply that the phase portrait is very different from that obtained in the limit $z_0
\to  \infty$. In the next section, we therefore start from reviewing the reference $z_0=\infty$ case.

\section{Trajectories for infinitely distant walls}\label{infinitely}

We describe briefly the seminal results in Ref.~\cite{ujhd} pertaining the relative motion of two spheres in unbounded shear flow, as determined by the equation of motion $\dot{\bm r}=\bm V^0(\bm r)$. The trajectories can be obtained from the two integrals $R_2$ and $R_3$ (notice a different labelling of the axes compared to Ref.~\cite{ujhd}. We have $y$ and $z$, and correspondingly $R_2$ and $R_3$, switched)
\begin{eqnarray}&&\!\!\!\!\!\!\!\!\!\!\!\!\!\!\!\!
R_2=y\exp\left(\int_r^{\infty}\frac{B(r')-A(r')}{1-A(r')}\frac{dr'}{r'}\right);\ \
R_3^2
%
=z^2\exp\left(2\int_r^{\infty}\frac{B(r')-A(r')}{1-A(r')}\frac{dr'}{r'}\right)\nonumber\\&&\!\!\!\!\!\!\!\!\!\!\!\!\!\!\!\!
\!-\!\int_r^{\infty}\!\!\!\frac{B(r')r'dr'}{1\!-\!A(r')}\exp\left(2\int_{r'}^{\infty}\frac{B(r'')\!-\!A(r'')}{1\!-\!A(r'')}\frac{dr''}{r''}\right).
\end{eqnarray}
We consider trajectories in the symmetry $xz$-plane ($y=0$) where $R_2=0$. The trajectories are given in the form $z=z(r)$ where ($r^2=x^2+z^2$),
\begin{eqnarray}&&\!\!\!\!\!\!\!\!\!\!\!\!\!\!\!\!
z^2(r)=
R_3^2\exp\left(2\int_r^{\infty}\frac{A(r')-B(r')}{1-A(r')}\frac{dr'}{r'}\right)
\!+\!\int_r^{\infty}\!\!\!\frac{B(r')r'dr'}{1\!-\!A(r')}\exp\left(2\int_r^{r'}\frac{A(r'')\!-\!B(r'')}{1\!-\!A(r'')}\frac{dr''}{r''}\right).
\end{eqnarray}
There are two types of trajectories: open and closed trajectories corresponding to $R_3^2>0$ and $R_3^2<0$, respectively. The regions in phase space occupied by open and closed trajectories are separated by the separatrix $z^s(r)$ whose equation is found by setting $R_3=0$,
\begin{eqnarray}&&\!\!\!\!\!\!\!\!\!\!\!\!\!\!
(z^s)^2\!=\!\!\!\int_r^{\infty}\!\!\!\frac{B(r')r'dr'}{1\!-\!A(r')}\exp\left(\!2\!\!\int_r^{r'}\!\!\!\frac{A(r'')\!-\!B(r'')}{1\!-\!A(r'')}\frac{dr''}{r''}\right).\label{separ}
\end{eqnarray}
We can obtain $z^s(r)$ at large $r$ using Eq.~(\ref{largr}),
\begin{eqnarray}&&\!\!\!\!\!\!\!\!\!\!\!\!\!\!\!\!
\int_r^{r'}\frac{A(r'')\!-\!B(r'')}{1\!-\!A(r'')}\frac{dr''}{r''}\approx \int_r^{r'}\frac{5dr''}{r''^4}=\frac{5}{3}\left(\frac{1}{r^3}-\frac{1}{r'^3}\right).\nonumber
\end{eqnarray}
The separatrix equation becomes (this asymptotic form was not presented in \cite{ujhd}),
\begin{eqnarray}&&\!\!\!\!\!\!\!\!\!\!\!\!\!\!\!\!
(z^s)^2\approx \exp\left(\frac{10}{3r^3}\right)\int_r^{\infty}\frac{16 dr'}{3r'^4}\exp\left(-\frac{10}{3r'^3}\right)
=\frac{8}{15}\left(\exp\left(\frac{10}{3r^3}\right)-1\right)\approx \frac{16}{9r^3}\approx  \frac{16}{9x^3}, \label{bsgd}
\end{eqnarray}
which shows that the separatrix asymptotically approaches the $x$-axis \cite{lin}. The surface obtained by rotation of this curve around the $z$-axis separates closed and open three-dimensional trajectories. The volume of closed trajectories is infinite due to divergence of two dimensional integral of $(x^2+y^2)^{-3/2}$.

We could not obtain a description of the particle-pair motion by integrals similar to $R_2$ and $R_3$ in the presence of the walls. For some trajectories, however, the wall is a small perturbation so that $\bm V(\bm r)\approx \bm V^0(\bm r)$ holds everywhere along the trajectory. The trajectory equation is then $\bm x(t)=\bm x^0(t)+\delta \bm x(t)$ where $\bm x^0(t)$ is a BG trajectory and $\delta \bm x(t)$  represents just a small modification. An example of these trajectories is the trajectory $a$ in Fig.~\ref{comparison}. These trajectories can be described with integrals of motion $R_i=R_i^0+\delta R_i$ where $\delta R_i$ is a small perturbation of the functional form of the $R_i$ due to the wall. This perturbation can be found from perturbation theory.
However this is of limited use since we are interested in trajectories for which the wall contribution is not small.

\section{Trajectories for a wall at finite distance}\label{finitely}

Here, we present the results of numerical simulations of the evolution equation of the inter-particle distance obtained in Sec. \ref{dista}. We apply the algorithm proposed in \cite{Filippov2000}, which allows us to
compute the hydrodynamic interactions in a system of $N$ spheres in a creeping flow. The algorithm is based on the multipole expansion of the Lamb solution for the fluid velocity field. We applied it to describe the motion in a system of two force- and torque-free solid spheres of unit radius in a shear flow for different distances $r$ between
the centers. Namely, for given components of the shear flow field and the vector
$\bm r$ connecting the sphere centers, we compute the velocity $\bm V$ in Eq.~(\ref{vls}).
Thus we determined the functions $A$, $B$, $K$, $L$ and $M$ for $r\geq 2.01$ using
the formulas in Ref.~\cite{ujhd}. When $r$ approaches the value $r=2$ the algorithm requires  a very large number of spherical harmonics into the solution expansion, which leads to a very large system of linear equations for the coefficients of the harmonics.

The functions $A$ and $B$ were therefore smoothly continued to $r=2$ using the asymptotic forms for almost touching spheres given by
\begin{eqnarray}&&\!\!\!\!\!\!\!\!\!\!\!\!\!\!\!\!
A(r)=1-4.077 (r-2)+O((r-2)^{3/2}),\ \
B\approx 0.406-\frac{0.78}{\ln \left[(r-2)^{-1}\right]}.
\label{tou}
\end{eqnarray}
The derivative of $B$ diverges at $r=2$, while the functions $A$, $K$, $L$ and $M$ are finite in the limit of touching spheres, $r\to 2$ and can be continued from $r\geq 2.01$ to $r<2.01$ using a linear Taylor series approximation. A similar approach was used in Ref.~\cite{arp}, where, however, continuation was used only below $2.0002$. Our main interest is in the behavior at larger $r$ so we did not undertake the detailed solution for the small values of $r-2$. A higher resolution is needed for the precise evaluation of the impact of the wall on the nearly touching BG trajectories and is left for future work.

Here, the equations of motion are generated employing the velocities given by the contributions (\ref{unpert},\ref{fk}). These equations
are solved numerically using the custom code in {\it Mathematica}, which reduces the integration step when the trajectory approaches the
vicinity of $r = 2$. In this region, the different trajectories are very close to each other and one has to resolve them accurately. This necessity is obvious already from the BG trajectories in the symmetry plane. All trajectories when the spheres pass in close vicinity to $r=2$ are closed. In other words, the trajectories that cross the $z$-axis at $z$ obeying $2\leq z\leq 2+\Delta$ are closed; however, those crossing at $z>2+\Delta$ are open where $\Delta$ is a small number. The quantity $\Delta$ obeys the equation
\begin{eqnarray}&&\!\!\!\!\!\!\!\!\!\!
(2+\Delta)^2\!=\int_{2+\Delta}^{\infty}\!\!\!\frac{B(r')r'dr'}{1\!-\!A(r')}\exp\left(2\int_{2+\Delta}^{r'}\frac{A(r'')\!-\!B(r'')}{1\!-\!A(r'')}\frac{dr''}{r''}\right),
\end{eqnarray}
as readily seen from Eq.~(\ref{separ}). The evaluation of $\Delta$ from this equation (not done in \cite{ujhd}) is beyond our scope here. Note however that Ref.~\cite{arp} provide $\Delta\sim 10^{-5}$.

\begin{figure*}
\begin{tabular}{ccc}
\includegraphics[width=0.4\textwidth]{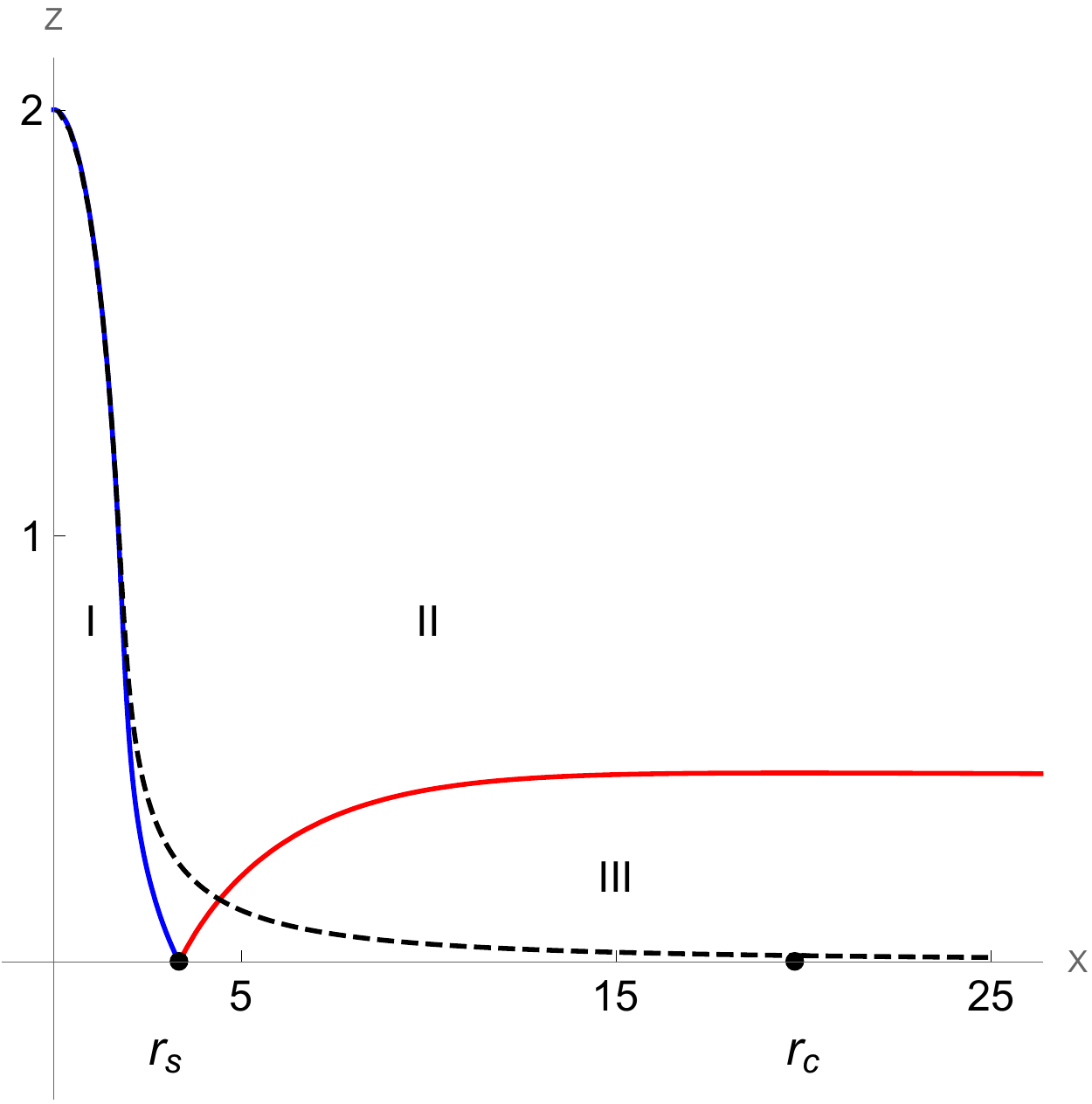} \hspace{1cm} &
\includegraphics[width=0.4\textwidth]{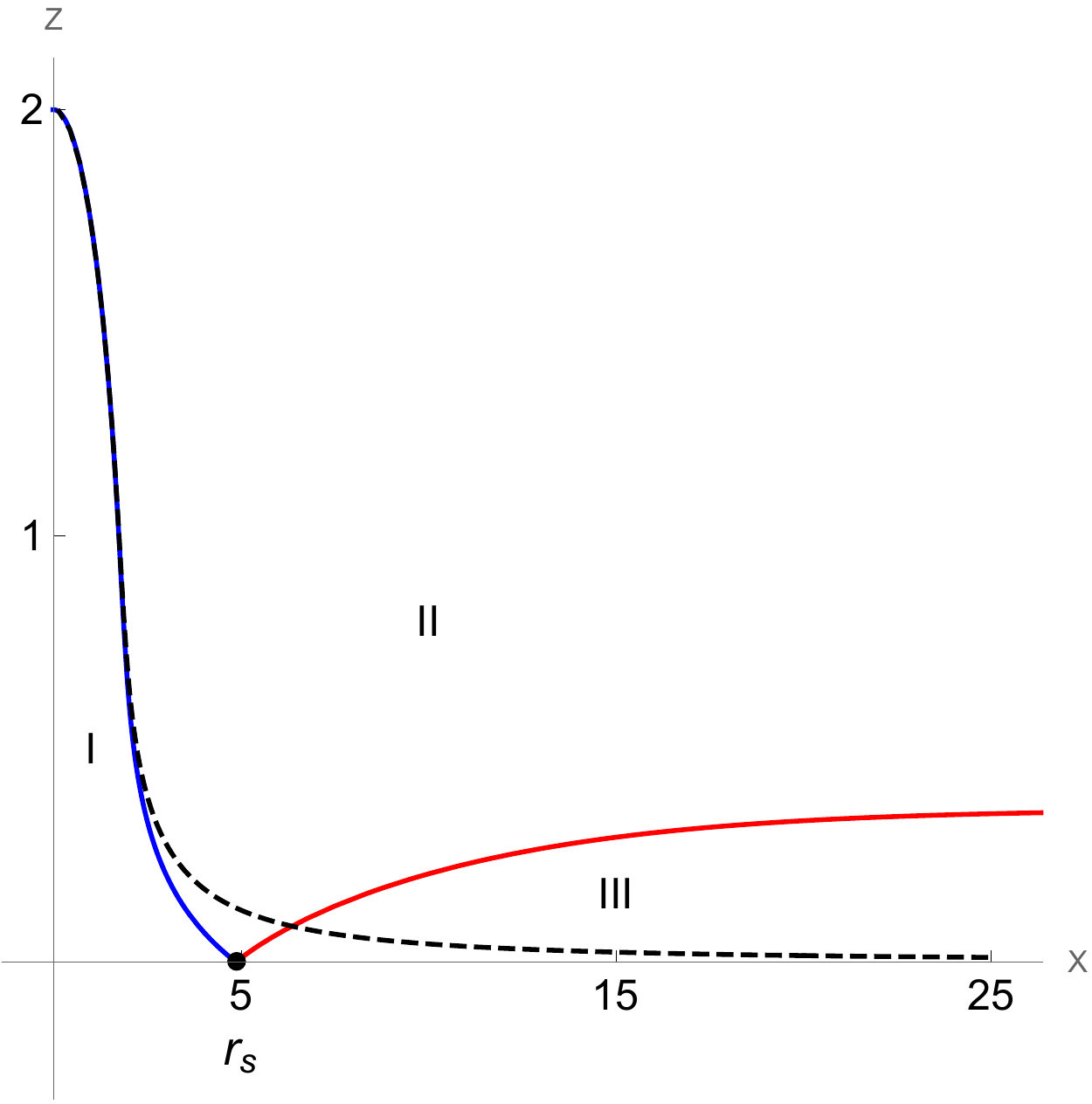}
\\
(a) & (b)
\end{tabular}
\caption{The phase portrait in the $xz$-plane for $z_0=5$ (a) and,
$z_0=10$ (b). The BG separatrix obeying $z^2=16/(9x^3)$ at large $x$ is depicted by the dashed (grey) line. For any finite $z_0$, the phase portrait contains two disconnected regions of closed trajectories, in contrast to one region at $z_0=\infty$. Region I, where all trajectories are closed and the spheres are close to each other, is similar to that at $z_0=\infty$. Region II is also similar to the $z_0=\infty$ case: all the trajectories are open and the vertical separation after the interaction returns to its original value.
Region III has no counterpart at $z_0=\infty$. This region contains both closed and open trajectories (see Fig.~\ref{Separatrix2D} for a more detailed description). The trajectories passing not far from the stationary point $r_c$ are closed, orbiting around this point. The swapping open trajectories instead are characterized by a sign reversal of the vertical component of the separation vector after the encounter. The region of swapping trajectories is bounded from one
side by the closed trajectories around $r_c$ and from the other side by open non-swapping trajectories.}
\label{fig:Fig4a}
\end{figure*}

The smallness of $\Delta$ implies that small perturbations can readily turn closed trajectory into an open one, which is indeed what the distant wall does as shown
in Fig.~\ref{fig:Fig4a}. The resolution of these small-scale effects demands high numerical precision.

To construct the separatrices (defined here as curves separating regions of qualitatively different behavior) in the $xz$-plane for given value of $z_0$ we first
find the critical point on the $x$-axis $(r_s,0,0)$ where the approximate value of $r_s$ is given in Section III.
One separatrix (red curve in Fig. \ref{fig:Fig4a}) is stable, see Fig.~\ref{fig:Schematics} and thus is computed
using integration of the original equations. The other separatrix (blue curve in Fig. \ref{fig:Fig4a}) is unstable as seen from Fig. \ref{fig:Schematics}. Thus, it is
found by
backward integration in time, for which it is stable, until the trajectory reaches the $z$-axis. All the trajectories below the blue curve (region I) are closed, while
those between the red and blue curves (region II) are open -- they correspond to
non-swapping trajectories. The trajectories between the red curve and the $x$-axis (region III)
can be divided into two classes -- open swapping trajectories (brown, black curves in the inset of Fig.~\ref{Separatrix2D}) and closed trajectories characterized by a
very large separation between the spheres (green, blue curves), see the captions of the Figures and detailed theory in the next Section.

\begin{figure}[h!]
\includegraphics[width=0.4\textwidth]{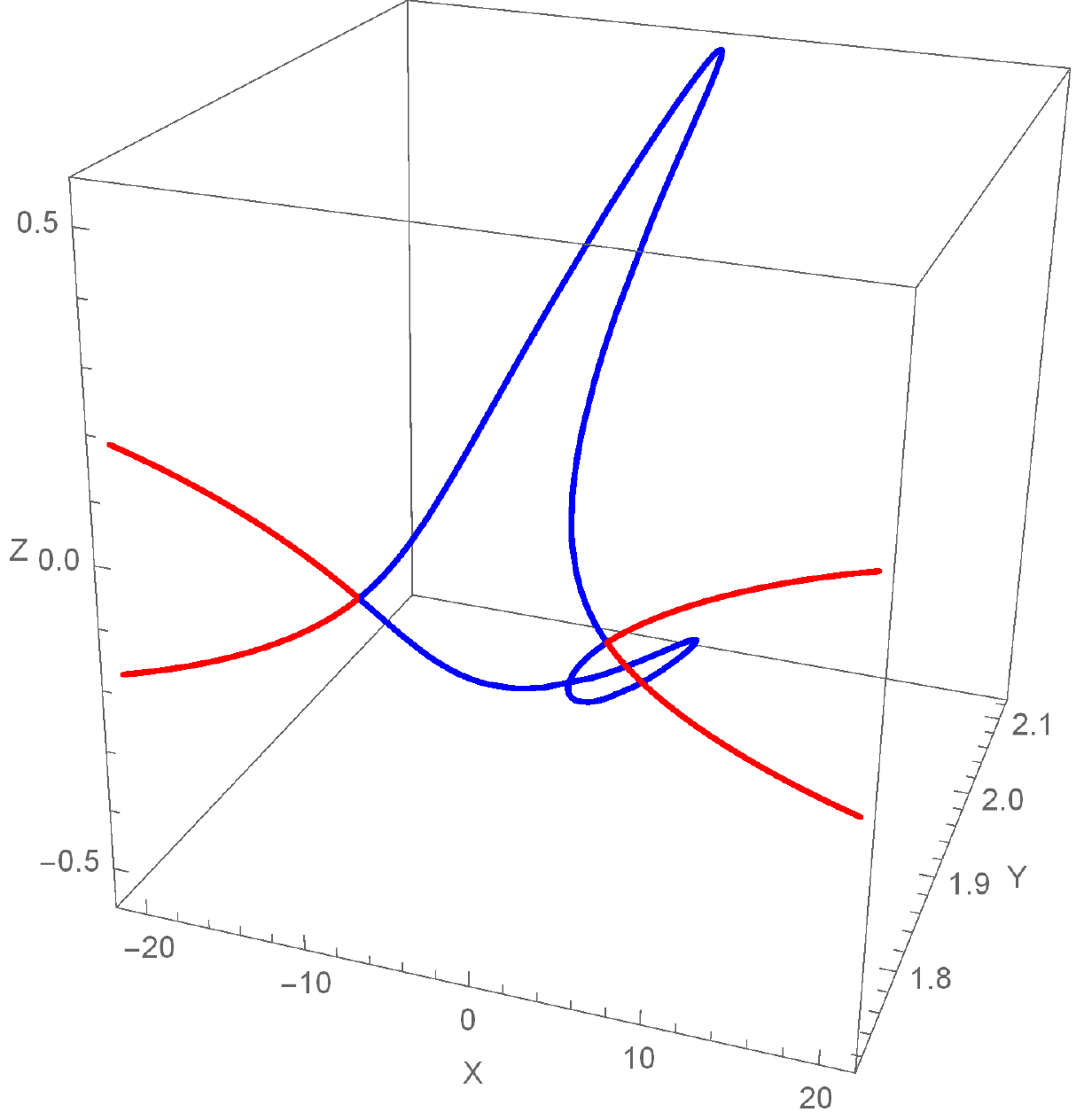}
\caption{The separatrices for $z_0=20$ corresponding to the
initial point $(r_s \cos\phi,r_s\sin\phi,0)$ with $\phi = 7\pi/90$.}
\label{Separatrix3D}
\end{figure}

\begin{figure}[h!]
\includegraphics[width=0.4\textwidth]{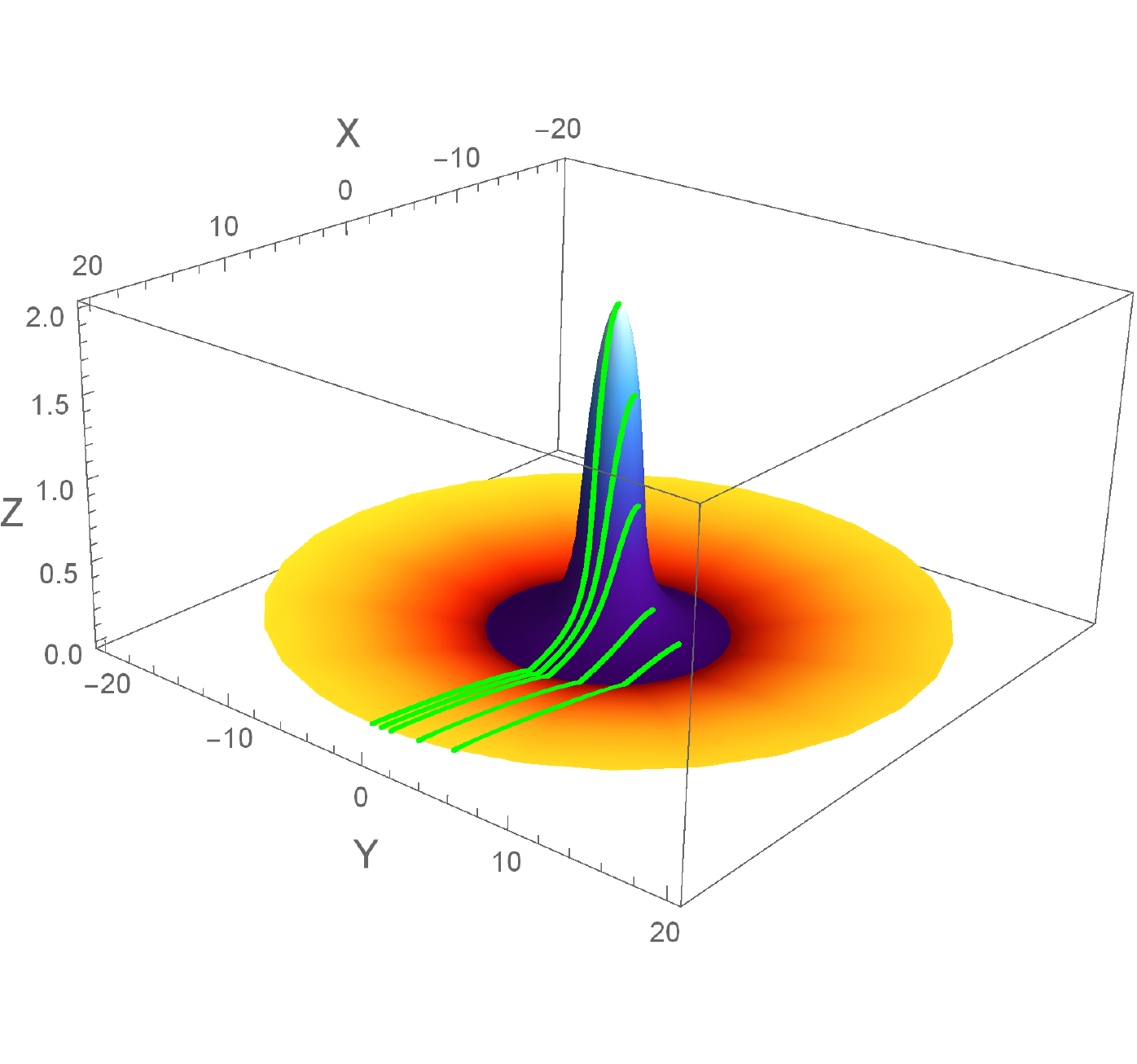}
\caption{Surface of rotation formed by the separatrices that pass through  $(r_s \cos\phi, \sin\phi,0)$ with $0\leq \phi\leq 2\pi$ at $z_0=20$. The green curves represent separatrices corresponding to $\phi = 0,\pi/36,\pi/18,5\pi/36,\pi/4$. The trajectories inside the region formed by the blue surface are closed BG-type orbits, whereas the trajectories inside the orange surface are either open swapping or closed dancing trajectories, as in region III in Fig.~\ref{Separatrix2D}, cf.\ Fig.~\ref{Separatrix3D}.}
\label{Separatrix3D_polar}
\end{figure}

We next consider three-dimensional trajectories. The axial symmetry of the governing equations
(\ref{rho})-(\ref{zz})
implies that the saddle points reside in the $xy-$plane on a circle with radius $r_s$.
For each point on this curve one can construct the corresponding separatrices in 3D
(see Fig.~\ref{Separatrix3D}, where the third neutral direction is given by the circle $r=r_c$, not shown).
All the separatrices belong to some surface of rotation (Fig.~\ref{Separatrix3D_polar}) which is obtained by the rotation of the curves in Fig.~\ref{Separatrix2D} around the $z$-axis.

\begin{figure*}
\begin{tabular}{cc}
\includegraphics[width=0.4\textwidth]{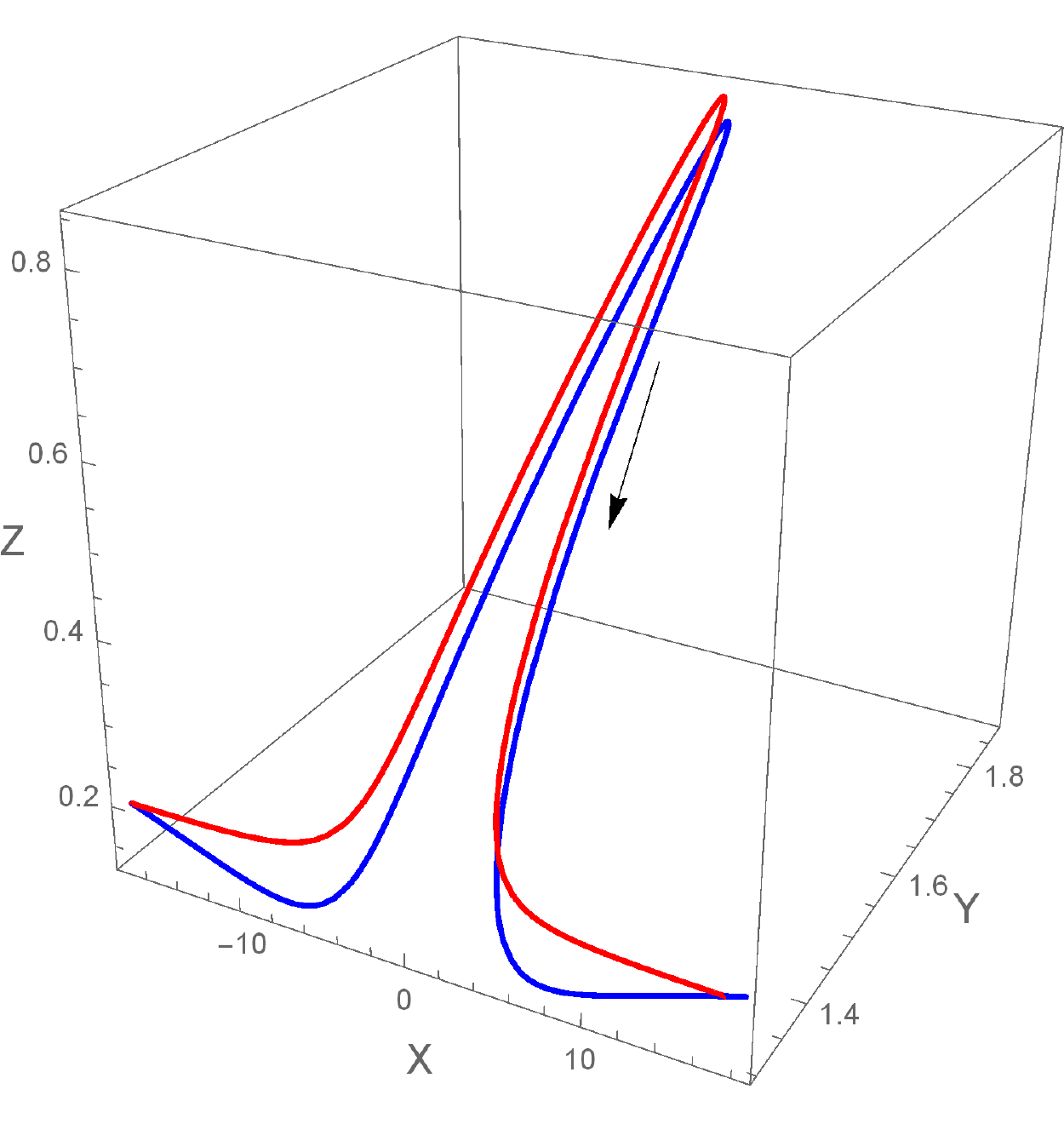} &
\includegraphics[width=0.4\textwidth]{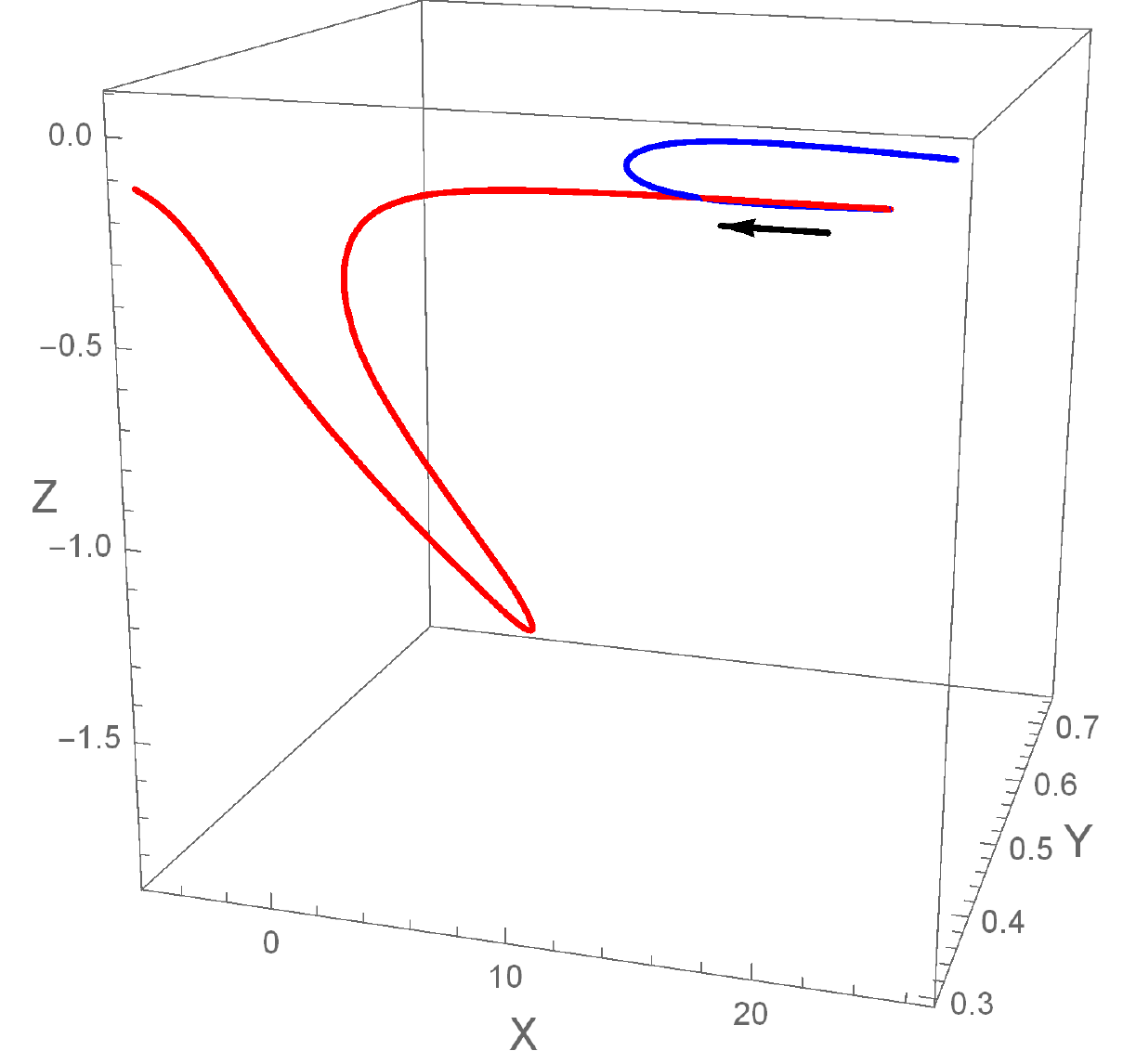}
\\
(a) & (b)
\end{tabular}
\caption{Comparison of representative open trajectories in the BG case (red) and in the case of a wall at a finite distance $z_0$ (blue): (a) typical open trajectories are qualitatively similar in both cases  ($z_0=20$); (b) for some initial conditions the presence of the wall results in the appearance of a swapping trajectory  ($z_0=5$). The black arrows indicate the direction of motion.}
\label{comparison}
\end{figure*}

At this point, it is instructive to compare the evolutions of the same representative initial conditions for $z_0=\infty$ and finite $z_0$.
The evolution of conditions that produce closed BG trajectories with small $x^2+z^2$ in the limit $z_0=\infty$ is only weakly influenced by far wall (unless passing near the BG separatrix where small perturbations are relevant), as in Fig.~\ref{FigX_BG_closed_trajectory1}, see the caption. In contrast, the trajectories with large $x^2+z^2$ may be very different as shown in Fig.~\ref{Fig8b_CircleRc} where the wall changes the evolution from an open trajectory to a closed one. The evolution of initial conditions leading to open trajectories for $z_0=\infty$
may be only slightly changed by the wall, as in Fig.~\ref{comparison}(a), or result in swapping as for the case in Fig.~\ref{comparison}(b).

\section{Theory of dancing-swapping region}\label{dancing-swapping}

\begin{figure}[h!]
\begin{center}
\includegraphics[width=0.4\textwidth]{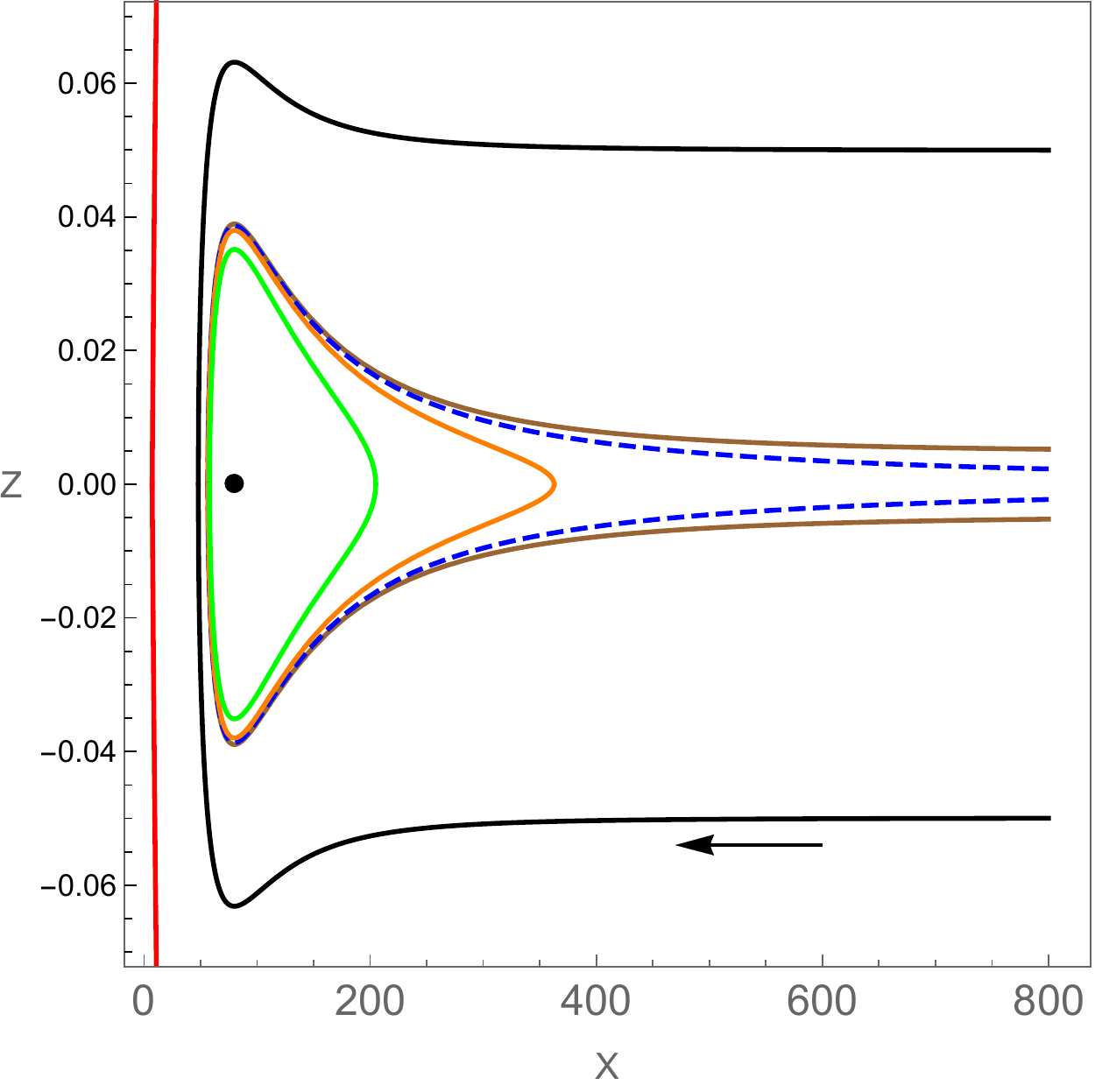}
\end{center}
\caption{The dancing-swapping region III of Figs. \ref{fig:Fig4a} for $z_0=20$. The red line is the region's boundary that crosses the $x$-axis at $\left(32 z_0^3/15\right)^{1/5}$. The dashed blue line separates open swapping and closed dancing trajectories and crosses the $x$-axis at $x_s=2\sqrt{2}z_0$. The black dot is the equilibrium point $(r_c=4z_0, 0)$.}
\label{Separatrix2D}
\end{figure}

In this Section we analyze the trajectories in the dancing-swapping region III, as shown in Figs.~\ref{fig:Fig4a}. We restrict the consideration to the symmetry plane $y=0$. All trajectories in this region cross the $x$-axis. These trajectories are of two types, both are qualitatively different from the BG theory. The swapping trajectories are open, each crosses the $x$-axis at a single point $x$ obeying $r_s<x<x_s$ where $r_s=\left(32 z_0^3/15\right)^{1/5}$ and $x_s=2\sqrt{2}z_0$ is determined below. For these trajectories the difference of the $z$ coordinates of the particles changes sign as a result of the hydrodynamic encounter (as for black curve in Fig.~\ref{Separatrix2D}. This sign-reversal corresponds to swapping of the vertical coordinates, see the Introduction. The larger crossing coordinate is, starting from $x=r_s$, the closer the trajectory is to the $x$-axis at large $x$. For the unique trajectory passing through $x=x_s$ the trajectory asymptotically approaches the $x$-axis indefinitely similarly to the BG's separatrix, dividing regions of open and closed trajectories. Finally, the trajectories that pass through a point $(x>x_s, 0, 0)$ are closed, each crossing the $x$-axis at two locations.

First we observe that the evolution of trajectories in the dancing-swapping region III admits $r\gg 1$ and thus can be simplified. It is readily seen numerically that, at least for $z_0\geq 5$ that are of interest here, we have $V^0_x(\bm r)\approx z$ within less than 15\% accuracy, meaning that the BG velocity difference is fully determined by the undisturbed shear flow. This is because the hydrodynamic interactions' correction to $V^0_x(\bm r)$ decays quickly with the spheres' separation, see Eqs.~(\ref{practical}), (\ref{largr}). We find from Eq.~(\ref{practical}) that we can use $V_x(\bm r)\approx z$ everywhere in region III. Moreover, we observe that $\delta V_z(x, 0, z)-\delta V_z(x, 0, 0)$ grows quadratically with $z$, see Appendix \ref{cartesian}. It is then found that since small-$z$ approximation holds (see below) then we can then use $\delta V_z(x, 0, z)\approx \delta V_z(x, 0, 0)$ in the whole region III. Finally, we can use the reduced Eq.~(\ref{velsc}). We find that the evolution of the trajectories in region III can be accurately described by the reduced system of equations,
\begin{eqnarray}&&\!\!\!\!\!\!\!\!\!\!\!\!\!\!\!\!
\dot x=z,\ \ \dot z=-\frac{8x }{3r^5}\!-\!\frac{5 xz^2}{r^5}+\frac{10x z_0^2(16z_0^2\!-\!x^2)}{(x^2+4z_0^2)^{7/2}},
\label{ds1}
\end{eqnarray}
where we assumed $z_0\geq 5$ and used Eq.~(\ref{bgz}). Furthermore, since region III is characterized by small $z$ then it is seen that $r\approx x$ and the second term in the RHS of the equation on $z$ can be dropped. Indeed, the ratio $15 z^2/8$ of this term to the first term in the RHS is small at moderate $x$ and not so small at larger $x$. However, at larger $x$ the time-derivative $\dot z$ is determined by the third term. Thus the second term is uniformly small everywhere in III as we verified numerically, and Eq.~(\ref{ds1}) is rewritten as
\begin{equation}
\dot x = z,
\quad
\dot z=-\frac{8}{3x^4}+
\frac{10x z_0^2(16z_0^2-x^2)}{(x^2+4z_0^2)^{7/2}}.
\label{ds2}
\end{equation}
The trajectories produced by this system in region III are indistinguishable from those produced by the full $\bm V(\bm r)$. The critical points of this approximate evolution obviously coincide with those obtained in Eq.~(\ref{critical_pts_eq}), so that, e. g. $10r_s z_0^2(16z_0^2-r_s^2)/(r_s^2+4z_0^2)^{7/2}=8/(3r_s^4)$ with $r_s= \left(32 z_0^3/15\right)^{1/5}$. Eliminating time variable we arrive at
\begin{equation}
\frac{d}{dx}z^2=-\frac{16}{3x^4}+
\frac{20x z_0^2(16z_0^2-x^2)}{(x^2+4z_0^2)^{7/2}}.
\label{ds3}
\end{equation}
The trajectory that crosses the $x$-axis at $x=x_i$ is given by the solution of the above equation and it reads
\begin{equation}
z^2=
\frac{16}{27x^3}
+\frac{20z_0^2(x^2-8z_0^2)}{3(x^2+4z_0^2)^{5/2}}
-\frac{16}{27x_i^3}
-\frac{20z_0^2(x_i^2-8z_0^2)}{3(x_i^2+4z_0^2)^{5/2}}.
\label{ds4}
\end{equation}
Setting here $x_i=r_s$ and using the condition on $r_s$ provided after Eq.~(\ref{ds2}), we arrive at the equation of the separatrix bounding the dancing-swapping region and separating it from region II (red line in Fig. \ref{fig:Fig4a})
\begin{equation}
z_{II, III}^2=
\frac{16}{27 x^3}
+\frac{20z_0^2(x^2-8z_0^2)}{3(x^2+4z_0^2)^{5/2}}+\frac{40z_0^2\left(48z_0^4-2r_s^2 z_0^2-r_s^4\right)}{9(r_s^2+4z_0^2)^{7/2}},\ \ r_s= \left(\frac{32 z_0^3}{15}\right)^{1/5}.
\label{ds5}
\end{equation}
We further find the following asymptotic behavior
\begin{equation}
z_{II, III}^2(x=\infty) =\frac{40z_0^2\left(48z_0^4-2r_s^2 z_0^2-r_s^4\right)}{9(r_s^2+4z_0^2)^{7/2}}\approx \frac{5}{3z_0},
\end{equation}
where the first equality holds down to $z_0=5$ and the last equality assumes $z_0\gg 1$. The equations confirm that region III has a finite width in $z$-direction, the fact underlying the validity of $\delta V_z(x, 0, z)\approx \delta V_z(x, 0, 0)$. The last equality provides the scaling law of growth of region III as the proximity to the wall decreases from $z_0=\infty$ to some finite value.

There is a unique value of $x_i=x_s$ for which the last two terms in Eq.~(\ref{ds4}) vanish and the trajectory asymptotes the $x$-axis at large $x$. This value is determined by the condition  $x_s^3=4(x_s^2+4z_0^2)^{5/2}/(45 z_0^2(8z_0^2-x_s^2))$. The solution is $x_s^2=8z_0^2-\epsilon$ with $\epsilon\approx 8 \sqrt{3}/(5\sqrt{2})$. The corresponding trajectory $z_{sw}$ is the separatrix of swapping and dancing trajectories,
\begin{equation}
z_{sw}^2=
\frac{16}{27x^3}
+\frac{20z_0^2(x^2-8z_0^2)}{3(x^2+4z_0^2)^{5/2}};\ \ \ z_{sw}^2(x\approx 2\sqrt{2} z_0)=0.
\end{equation}
This asymptotic behavior of this separatrix at $x\gg z_0$ is $z_{sw}^2\sim 20z_0^2/(3x^3)$. Remarkably this is the same behavior as the BG asymptote given by Eq.~(\ref{bsgd}), however with a much larger coefficient. Since the three-dimensional separatrix is obtained by revolution around the $z$-axis, we conclude that the volume of closed dancing trajectories is infinite. Thus the wall does not regularize the divergences in the stress calculation at the second order in concentration of \cite{bgst}. The volume of swapping trajectories is also infinite. Finally we remark that long-distance behavior of trajectories in regions I and II can also be described using the approach of this Section, however the global behavior in those regions involves close positions of the spheres and demands the full formulas.

\section{Direct numerical simulation of a particle pair in Poiseuille flows}\label{dns}

In this Section we provide evidence of the existence of the neutral equilibrium point $(r_c, 0, 0)$ from direct numerical simulations of the motion of a pair of particles in the Poiseuille flow. We simulate the Navier-Stokes equations with appropriate boundary conditions for Reynolds number of $0.1$. We consider a moderate distance from the wall, of $z_0=5$, demonstrating that the theory is accurate in this configuration. In this way, we provide confirmation of the theory and demonstrate that the theory holds down to rather small $z_0$.

We use interface-resolved, direct numerical simulations. The particles are simulated either as solid spheres using an immersed boundary method or as liquid droplets using the interface-correction level set/ghost fluid method; see Refs.\ \cite{ibm,icls,flow-assist} and the Appendix B of \cite{2017}, for detailed descriptions of the governing equations and their numerical treatments.

Fig.\ \ref{fig:setup} illustrates the schematic of the simulation setup. Here, two neutrally buoyant particles are transported inside a rectangular channel of dimensions $L_x$, $L_y$, and $L_z$, that are at least an order-of-magnitude larger than the particle radius $a$. The undisturbed flow is the Poiseuille flow shifted backwards by a constant velocity so that the position of the first particle remains roughly unchanged throughout the simulation \cite{mframe}.
The particle pair is initially placed adjacent to the bottom wall, with $z_0=5$ and $L_z=64$. $L_x$ and $L_y$ are chosen to be large enough so that the imposed boundary conditions (periodic or inflow/outflow) do not qualitatively affect the particle motion, which we verified by checking that changes in $L_x$ and $L_y$ do not affect the results appreciably. Thus we used $L_x=12$ and $24$, and increased $L_y$ from $60$ up to $80$.

\begin{figure}[t]
\includegraphics[width=0.4\textwidth]{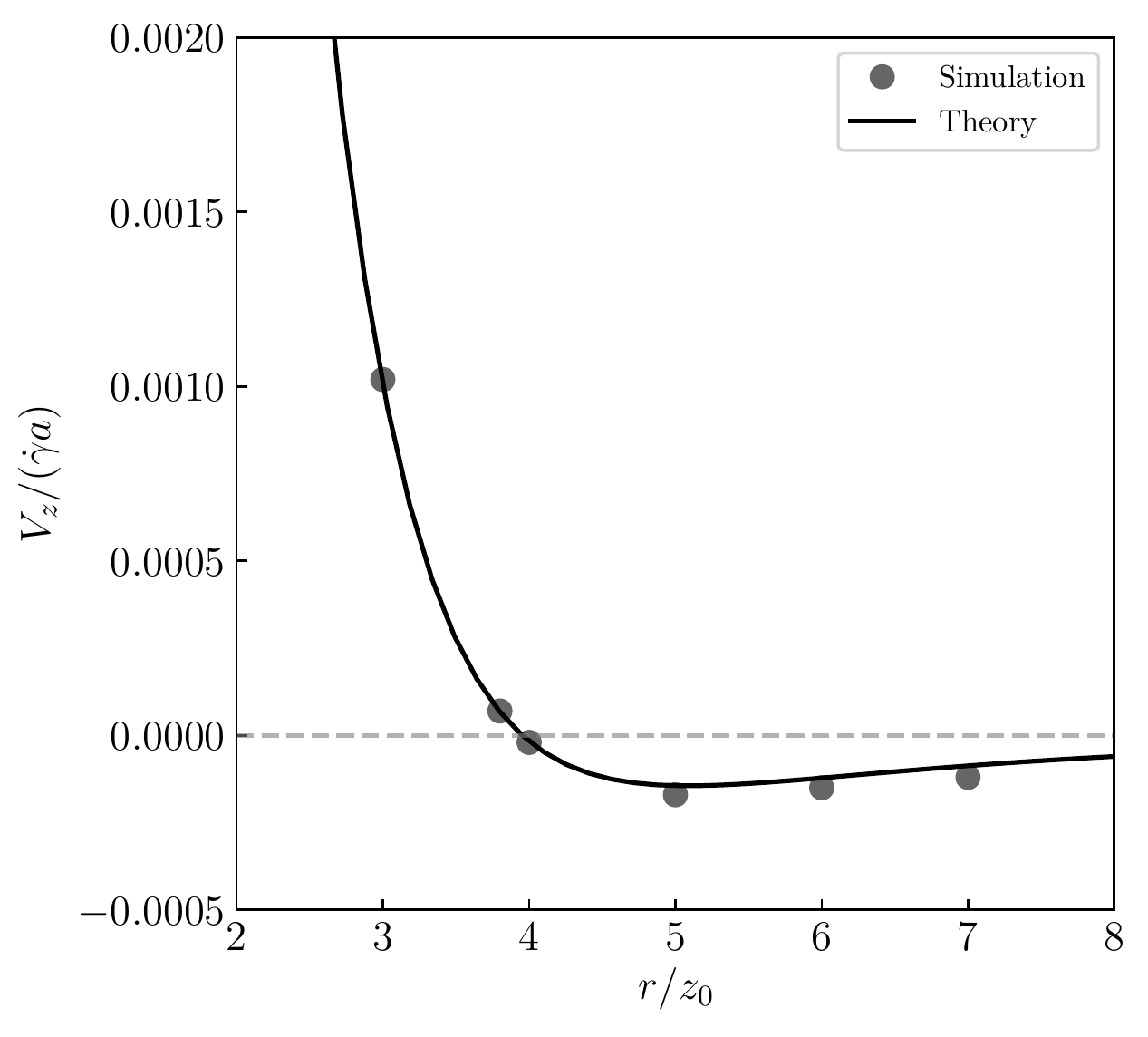}
\caption{Asymptotic vertical velocity of particle 2 relative to particle 1 as a function of the horizontal separation, at $z_0=5a$ (cf.\ Fig.\ \ref{fig:setup}). We focus
 on the range where the theory predicts change of sign of the velocity and the associated critical point. The theory is seen to hold accurate predictions even in geometrically confined Poiseuille flows.}
\label{fig:vel}
\end{figure}

Fig.\ \ref{fig:vel} depicts the vertical component of the relative velocity $V_z$ of two solid particles at various initial separations $r/z_0$, obtained asymptotically upon their release. That is, we extract $V_z$ from the simulations when both particles are still approximately at the same vertical position $z_0$ within the accuracy of $10^{-4}$. The theoretical values are computed according to Eqs.\ \eqref{bgz} and \eqref{fok}, which apply since the minimal considered distance is $15$.
Remarkably, we observe a close agreement between the theoretical prediction and the numerical simulation, from the smallest studied distance of $r=3z_0$.
This is despite that the simulations are performed in a pressure-driven channel flow in the presence of two walls and $z_0$ is not so large.
The deviation of the numerical results from the theoretical values at $r/z_0 \geq 5$ is probably due to numerical confinement; as the particles are further separated, larger computational boxes would be necessary to accurately isolate the interaction due solely to the neighbouring particle.

\section{Conclusions}

We presented here the theory of the hydrodynamic interactions of two spheres in a shear flow in the presence of a plane rigid wall. This theory provides a reference for consistent direct numerical or experimental studies of the particles' trajectories. Some of the predictions of the theory have been confirmed by direct numerical simulations in Poiseuille flow, demonstrating that neglecting the farthest wall is a valid assumption and the theory holds at least down to distances from the wall of 5 particle radii, $z_0 \approx 5a$.

The immediate use of our work is the determination of the limitations of the BG theory \cite{ujhd}, see also \cite{kim}. Our theory indicates that for inter-particle distances $r/a$ much smaller than $(z_0/a)^{3/5}$ variations of the inter-particle velocity with respect to the BG velocity is small.
This condition, $r\ll z_0^{3/5} a^{2/5}$, is stricter than the rough estimate $r \ll z_0$. If this condition is not fulfilled, the trajectories are significantly changed  in comparison with the BG predictions, both quantitatively and qualitatively.

However, corrections due to the wall can also be relevant at $r\ll z_0^{3/5} a^{2/5}$. The reason is that the global behavior of the BG trajectories passing near the separatrix is sensitive to small perturbations. Thus, perturbations originating from the wall presence, Brownian noise, gravity, finite roughness of the particles' surface or any other source, may easily change the global portrait of the interactions. All the closed trajectories of the classic BG solution \cite{ujhd} can be altered quite significantly by small perturbations, since they all pass near the separatrix. Indeed when particles, orbiting around each other in the symmetry plane, reach the vertical (side-by-side) orientation, the maximal distance between them is of order of $10^{-5} a$, see \cite{arp}. The wall, even a distant one, can produce a small upward displacement which would shrink the region of closed trajectories. This is in fact what we see in the simulations where the separatrix in the presence of the wall crosses the $z$-axis at shorter distance from $z-2 a$ than without the wall. However, our simulations are not built for resolving distances as small as $10^{-5} a$ so this initial observation demands further, more accurate studies, which can take advantage of the evolution equation for the inter-particle separation derived here.

It is of interest to discuss here how our findings affect the applicability of the BG theory in various practical situations. We consider as an example the linear approximation to the turbulent flow at scales much smaller than the viscous scale of turbulence $\ell_\eta$, \cite{frisch}. This approximation can be used for describing collisions of two small droplets in turbulence, relevant, among others, to the problem of rain formation \cite{clc}. The linear approximation breaks down at a finite distance $\ell_\eta$ from the colliding particles with radii $a\ll \ell_\eta$. Our work implies that at distances $r\sim \ell_\eta^{3/5} a^{ 2/5}$ the BG description may be inaccurate. Thus the applicability of the BG description in the region of significant hydrodynamic interactions, $r\sim a$, demands $(\ell_\eta/a)^{3/5}\gg 1$. We find that for strong cloud turbulence with energy dissipation rate of $2000$~cm$^2$/s$^3$ this condition might fail already for droplet sizes of $50$~$\mu$m. In contrast, if the naive criterion of $\ell_\eta/a\gg 1$ was used, then the approximation would still be valid. In this specific problem,  however, other difficulties that involve non-stationarity of the velocity gradients also appear. Thus $\bm V^0(\bm r)$ would hold with the instantaneous value of the coefficients of the linear flow considered by \cite{ujhd}. However the periods of revolution, as indicated
above, can be much larger than the inverse shear rate, which in turbulence determines the characteristic time of variations of the flow gradients. This would make the BG trajectories inapplicable. Another difficulty arises from the finiteness of the Reynolds number, which may be relevant already at $30$~$\mu$m, see, e.g., \cite{fi2015}. This would demand the introduction of corrections to the BG theory due to the non-linear term in the Navier-Stokes equations. These questions are of high interest due to
the ubiquitous occurrence of collisions of small particles in turbulence and are left for future work.

We notice that the problem considered here seemingly has a hidden symmetry. The presence of the wall makes the top-down symmetry, which is displayed by the trajectories, non-evident. Moreover, it is not so evident why the separatrices form a surface of rotation in both the BG and our cases.

Strictly speaking our analysis is not complete. The leading order correction that we found has naive order of smallness $z_0^{-2}$, and not the actually holding $z_0^{-3}$. We found this from the detailed calculation that revealed the vanishing of the  $z_0^{-2}$ contribution due to symmetry. It is possible that the symmetry would be irrelevant for the next order term which has the naive order of magnitude of $z_0^{-3}$ and it cannot be neglected. We consider this scenario implausible and make the conjecture that the next order term is actually $\mathcal{O}(z_0^{-4})$ and can be consistently neglected. Proving this conjecture theoretically is a formidable task which was not undertaken here. It seemed more practical to test the predictions that we made by direct numerical simulations of the motion of two spheres in a shear flow in the presence of a wall. The performed numerical simulations of the Poiseuille flow closely confirmed the predictions of our theory.

Another confirmation of our theory comes from the previous, unguided by the theory, simulations of \cite{agto}. This work considered the shear flow between two parallel planes induced by the motion of the upper plane. This problem, with both walls included, could be considered as in Sec.\ \ref{interpair} by using the Green's function for the Stokes flow between two infinite planes \cite{LironMochon}, which however is beyond the scope of the present paper.
The interacting spheres in \cite{agto}, however, were located closer to the immobile lower wall which makes our theory applicable at least qualitatively.
The phase portrait of \cite{agto} for the evolution of the inter-particle distance in the symmetry plane agrees remarkably well with that provided here, though it lacks the neutral equilibrium point and the closed trajectories revolving around it. The authors observed the saddle point at $z_0=4.8$ with distance $10$ between the walls. In this case, our theory applies only qualitatively. However, when we use our formula $(32 z_0^3/15)^{1/5}$ for the position of the saddle point, we find that our prediction agrees very well with the numerical findings of \cite{agto}. All these provide strong evidence for validity of our theory.

The complete proof showing that our open trajectories with sign-reversal of the vertical separation describe swapping of the vertical positions
requires the computation of the vertical coordinate of the center of mass after the interaction.
 Although as we argued above, this seems inevitable,  a proof demands the study of the motion of the center of mass, which was not undertaken here (the formulae of Sec.~\ref{interpair} can be used for this aim). For an unbounded shear flow, the motion of the center of mass could be obtained using the shear resistance matrix, function of
 the instantaneous distance between the spheres. This matrix can be written in terms of scalar coefficient functions, similar to $A$ and $B$, with the asymptotic form of this matrix  obtained at large separations in \cite{sh} (see also \cite{arp}). Considering this matrix
and the solution for the inter-particle distance as a function of time as given, one can readily find the center of mass velocity as a function of time. In our case the calculations are even more involved due to the presence of the wall. This is therefore left for future work.

The numerical and experimental tests of our predictions may focus on the emergence of the neutrally stable bound state, when the particle pair flows as a whole
at some fixed distance from the wall $z_0$. The horizontal component of the inter-particle distance in this state belongs to the circle of radius $4 z_0$, although at small $z_0$ some deviations from $4 z_0$ must occur.

Since the experiments of \cite{tab0,tabeling,flow-assist} used droplets and not rigid particles, we shall briefly address how the results obtained for the rigid particles here would change for droplets. Close interactions of rigid particles and droplets are quite different, both qualitatively and quantitatively, see e. g. \cite{kim}. However at large separations, when the effects of the wall are most relevant, the differences seem to be less significant. We have confirmed this again using direct numerical simulations for two liquid droplets in the same setup as for the solid particles in Sec.~\ref{dns}. We verified that $V_z>0$ at $r/z_0=3$ while $V_z\lesssim 0$ at $r/z_0=5$. Thus, there is a point within this range where the velocity vanishes, as in the case of rigid particles. Therefore at least the prediction of the stationary point holds also in the case of liquid droplets. This suggests that the existence of states of marginal equilibrium is a robust phenomenon for pair of particles flowing next to a wall.

The theory presented here has direct generalisations to other distant boundaries.  The developed approach can also be used to study the hydrodynamic interactions between suspended particles in other confined shearing flows, such as, e.g., Couette flow. The case of a third particle at a finite distance from the pair of spheres in an unbounded shear flow is also of interest. When the driving flow is enclosed between two parallel planes (i.e., a slit geometry), as in \cite{agto}, the inclusion of the second plane is required for a theoretical analysis, as suggested above.

The present finding of stable configurations of pairs of particles due to hydrodynamic interactions is probably due to the fact that the position of one of the three bodies in interaction - the wall, - is fixed. The question whether such configurations can exist for three, or a larger number of flowing particles is left for future work.

\section*{Acknowledgement}
The authors wish to thank K.~I.~Morozov for fruitful discussions. Z. Ge thanks P.~S.~Costa for technical discussions on the numerical solver. The work is supported by the Microflusa project. The Microflusa project receives funding from the European Union Horizon 2020 research and innovation programme
under Grant Agreement No. 664823.



\newpage
\begin{appendices}

\section{Integral representation of shear flow round spheres}\label{shear}

We derive here the integral representation of the flow round spheres driven by shear in an unbounded fluid. The flow obeys,
\begin{eqnarray}&&\!\!\!\!\!\!\!\!\!\!\!\!\!
\nabla p\!=\!\eta \nabla^2 \bm u,\ \ \bm u(\infty)\sim \dot{\gamma} z{\hat x},\ \ \nabla\cdot\bm u=0,\ \
\bm u(S_{\alpha})=\bm V_{\alpha}+\bm \Omega_{\alpha}\times (\bm x-\bm x_{\alpha}),
\label{rescaled}
\end{eqnarray}
where $\dot{\gamma}$ is the shear rate and as in the main text $\alpha=1, 2$ are the indices of the spheres and $x_{\alpha}$ are the coordinates of the centers. Translational and angular velocities are determined from the conditions that the fluid applies to each particle zero net force and torque,
\begin{eqnarray}&&\!\!\!\!\!\!\!\!\!\!\!\!\!
\int \sigma_{ik}dS_{\alpha k}=0,\ \ \int (\bm x-\bm x_{\alpha})\times \sigma_{ik}dS_{\alpha k}=0,\ \ \sigma_{ik}\equiv -p\delta_{ik}+\eta(\nabla_iu_k+\nabla_ku_i)
\end{eqnarray}
where $\sigma_{ik}$ is the stress tensor. We use the Lorentz-type identity for $\bm x'$ outside the volume of the spheres,
\begin{eqnarray}&&
 u_{i}(\bm x')\delta(\bm x'-\bm x)=\frac{\partial}{\partial x_k'}\left[\frac{Y_{il}(\bm x-\bm x')\sigma_{lk}(\bm x')}{8\pi \eta}+u_l(\bm x')\Sigma_{ilk}(\bm x- \bm x')\right].\label{fd}
\end{eqnarray}
where $Y_{il}$ is defined in Eq.~(\ref{inter}) and $\Sigma_{lik}$ defines the stress tensor of the Stokeslet. We have (our definition differs from \cite{kim} by insignificant permutation of indices of the symmetric tensor $\Sigma_{ilk}$),
\begin{eqnarray}&&\!\!\!\!\!\!\!\!\!\!
Y_{il}=\frac{\delta_{il}}{r}+\frac{r_ir_l}{r^3},\ \ \Sigma_{ilk}=\frac{1}{8\pi}\left(\frac{\partial Y_{il}}{\partial r_k}+\frac{\partial Y_{ik}}{\partial r_l}\right)-\frac{r_i\delta_{lk}}{4\pi r^3}=-\frac{3}{4\pi}\frac{r_ir_lr_k}{r^5},\ \
\frac{\partial}{\partial x'_k}\Sigma_{ilk}(\bm x- \bm x')=\delta_{il}\delta(\bm x-\bm x'). \label{free}
\end{eqnarray}
Integrating Eq.~(\ref{fd}) over $\bm x'$ outside the particles, we find
\begin{eqnarray}&&\!\!\!\!\!\!\!\!\!\!
u_i(\bm x)\!=\!\!\int_{S_{\infty}}\!\!\!\left(\frac{Y_{il}(\bm x\!-\!\bm x')\sigma_{lk}(\bm x')}{8\pi \eta}\!+\!u_l(\bm x')\Sigma_{ilk}(\bm x\!-\!\bm x')\right) dS'_k
\!-\!\sum_{\alpha}\int_{S_{\alpha}}\!\!\!\frac{Y_{il}(\bm x-\bm x')\sigma_{lk}(\bm x')dS_k'}{8\pi \eta},\label{in}
\end{eqnarray}
where the direction of the normals is outward from the surfaces and $S_{\infty}$ is the spherical surface with radius $R$ taken to infinity. The prime designates that the integrals are over the $\bm x'$ variable and the term with an integral of $\Sigma$ over the particle surfaces vanishes by the rigid body boundary condition \cite{kim,ps,2017}. We observe that the disturbance of the flow caused by the spheres vanishes at infinity so that,
\begin{eqnarray}&&\!\!\!\!\!\!\!\!\!\!\!\!\!\!\!\!
\bm u\!\sim\! \dot{\gamma} z{\hat x}\!+\!o(const),\ \ \sigma_{lk}\!\sim\! \eta\dot{\gamma}(\delta_{lx}\delta_{kz}\!+\!\delta_{kx}\delta_{lz})\!+\!o(r^{-1}). \nonumber
\end{eqnarray}
We use these asymptotic forms for obtaining the integrals over $S_{\infty}$. Keeping the lowest order non-vanishing term in the Taylor series of $Y_{il}(\bm x-\bm x')$ in $\bm x$,
\begin{eqnarray}&&\!\!\!\!\!\!\!\!\!\!\!\!
\int_{S_{\infty}}\!\!\frac{Y_{il}(\bm x\!-\!\bm x')\sigma_{lr}dS'}{8\pi \eta \dot{\gamma}}\!=-\!x_m\!\int_{S_{\infty}}\!\!(\delta_{lx}z'\!+\!x'\delta_{lz})
\frac{dS'}{8\pi R} \frac{\partial Y_{il}(\bm x')}{\partial x'_m}.\nonumber
\end{eqnarray}
We find using the form of $Y_{ik}$ in Eq.~(\ref{free}),
\begin{eqnarray}&&\!\!\!\!\!\!\!\!\!\!\!\!\!\!
\frac{\partial Y_{il}}{\partial r_m}=\frac{r^2(r_l\delta_{im}+r_i\delta_{lm}-r_m\delta_{il})-3r_ir_lr_m}{r^5},
\end{eqnarray}
and obtain,
\begin{eqnarray}&&\!\!\!\!\!\!\!\!\!\!\!\!
\int_{S_{\infty}}\!\!\frac{Y_{il}(\bm x\!-\!\bm x')\sigma_{lr}dS'}{8\pi \eta \dot{\gamma}}\!=\!-\frac{x_m}{2}\!\left\langle (\delta_{lx}z\!+\!x\delta_{lz})
\left(x_l\delta_{im}+x_i\delta_{lm}-x_m\delta_{il}-3x_ix_lx_m\right)\right\rangle,
\end{eqnarray}
where angular brackets stand for averages over the unit sphere,
\begin{eqnarray}&&\!\!\!\!\!\!\!\!\!\!
\left\langle x_ix_k\right\rangle=\int_{x=1} \frac{x_ix_k dS}{4\pi}=\frac{\delta_{ik}}{3},\ \
\left\langle x_ix_kx_lx_m\right\rangle=\frac{\delta_{ik}\delta_{lm}+\delta_{il}\delta_{km}+\delta_{im}\delta_{kl}}{15}.\label{average}
\end{eqnarray}
Further, by collecting the different terms,
\begin{eqnarray}&&\!\!\!\!\!\!\!\!\!\!\!\!
\int_{S_{\infty}}\!\!\frac{Y_{il}(\bm x\!-\!\bm x')\sigma_{lr}dS'}{8\pi \eta \dot{\gamma}}\!=\!\frac{\delta_{iz}x+\delta_{ix}z}{5}.
\end{eqnarray}
We consider similarly the remaining integral over $S_{\infty}$,
\begin{eqnarray}&&\!\!\!\!\!\!\!\!\!\!
\int_{S_{\infty}}\!\!u_l(\bm x')\Sigma_{lik}(\bm x- \bm x') dS'_k=\dot{\gamma}\int_{S_{\infty}}\!\!z'\Sigma_{xik}(\bm x- \bm x') dS'_k
=x_m\dot{\gamma}\int_{S_{\infty}}\!\!z'\frac{\partial \Sigma_{xik}(\bm x')}{\partial x'_m} dS'_k.
\end{eqnarray}
We have from Eq.~(\ref{free}) that,
\begin{eqnarray}&&\!\!\!\!\!\!\!\!\!\!
\frac{\partial \Sigma_{xik}}{\partial r_m}\!=\!\frac{3}{4\pi}\left(\frac{5x r_ir_kr_m}{r^7}\!-\!\frac{r_ir_k\delta_{mx}\!+\!r_ix\delta_{mk}\!+\!r_kx\delta_{mi}}{r^5}\right),\nonumber
\end{eqnarray}
and
\begin{eqnarray}&&\!\!\!\!\!\!\!\!\!\!
\int_{S_{\infty}}\!\!u_l(\bm x')\Sigma_{lik}(\bm x- \bm x') dS'_k=3x_m\dot{\gamma}\left\langle\left( 4x z r_ir_m -zr_i\delta_{mx}
-xz\delta_{mi}\right) \right  \rangle.
\end{eqnarray}
We obtain using Eq.~(\ref{average}),
\begin{eqnarray}&&\!\!\!\!\!\!\!\!\!\!
\int_{S_{\infty}}\!\!u_l(\bm x')\Sigma_{lik}(\bm x- \bm x') dS'_k=\dot{\gamma}\left(\frac{4\delta_{ix}z-\delta_{iz}x}{5}\right).
\end{eqnarray}
Collecting the terms in Eq.~(\ref{in}), we obtain the integral representation given by Eq.~(\ref{inter}) in the main text. Thais representation leads to Eq.~(\ref{intb}) in the main text when taking $\bm x$ on the surface of one of the spheres and using the proper boundary condition. The representation could also be derived by considering the standard integral representation \label{ps,kim} for the correction flow $\bm u-\dot{\gamma} z{\hat x}$, which also obeys the Stokes equation. In that approach one would need to evaluate integrals on the particles' surfaces instead of $S_{\infty}$ to find Eq.~(\ref{inter}).

The integral representation in Eq.~(\ref{inter}) gives readily the multipole expansion of the flow at large distances $x\gg x'$, see \cite{ps,kim}. The leading order term is provided in Eq.~(\ref{far}) of the main text
where we use the condition of zero force. This approximation holds at $|\bm x-\bm x_{\alpha}|$ much larger than the radii of the spheres.

\section{Dynamical equations in Cartesian coordinates}
\label{cartesian}

We can derive an explicit expression for $\delta V_i$ using Eqs.~(\ref{dipole}) and (\ref{tensors}). To this end, we introduce $s^2 = x^2+y^2+4z_0^2$,
$\sigma=z^2-4z_0^2$ and $g^2=x^4+y^4+4z^2 z_0^2$. We find after simplifications that $\delta V_i$ can be written as sums over four components $c_i^k$ (we use dimensionless time $\dot{\gamma} t$),
\begin{eqnarray}&&\!\!\!\!\!\!\!\!\!\!\!
\delta V_x\!=\! \frac{5z}{r^4}\!\sum_{k=1}^4\! c_x^k,
\ \
\delta V_y\! = \!\frac{5xyz}{r^4}\! \sum_{k=1}^4 \!c_y^k,
\ \
\delta V_z\! =\! \frac{5x}{r^4}\!\sum_{k=1}^4\! c_z^k.
\label{delV}
\end{eqnarray}
The components of $\delta V_x$ are given by,
\begin{eqnarray}&&\!\!\!\!\!\!
c_{x}^1\!  =\!
\frac{y^2(2s^2x^2\!-\!(s^2\!-\!5x^2)\sigma)(r^2L\!+\!2x^2M)}{2 s^7},\ \ c_{x}^3\! =\!x^2 c_{y}^3 ,
\ \ c_{x}^2\! =\!
\frac{2z_0}{\sigma^2}((1\!+\!K)r^4\!+\!(x^2\!+\!z^2)r^2L(r)\!+\!2x^2z^2M),
\nonumber \\&&\!\!\!\!\!\!
c_{x}^4 \!=\!-\frac{x^2\sigma}{2s^7}[r^2(5y^2\!-\!s^2)L\!+\!
(s^2(r^2\!+\!2x^2)\!-\!5g^2)M],
\end{eqnarray}
where the components of $\delta V_y$ read,
\begin{eqnarray}&&\!\!\!\!\!\!
c_{y}^1 \!=\!
\frac{(2s^2y^2\!-\!(s^2\!-\!5y^2)\sigma)(r^2L\!+\!2x^2M)}{2s^7},\ \ c_{y}^2 \!=\!
\frac{2z_0}{\sigma^2}(r^2L\!+\!2z^2M),\ \
c_{y}^3 \!=\! \frac{(s^2\!-\!3y^2)r^2L\!+\!(3g^2\!-\!r^2s^2)M}{3s^5},
\nonumber \\&&\!\!\!\!\!\!
c_{y}^4\! =\!
-\frac{\sigma}{2s^7}[r^2(5y^2\!-\!3s^2)L\!+\!
(s^2(r^2+2y^2)\!-\!5g^2)M].
\end{eqnarray}
Finally the components of $\delta V_z$ are,
\begin{eqnarray}&&\!\!\!\!\!\!
c_{z}^1\! =\!
\frac{\sigma}{2s^7}(s^2\!-\!20z_0^2)
[r^4(1+K+L)\!+\!2z^2(r^2\!-\!z^2)M],
\ \
c_{z}^2  \!=\!
\frac{2z^2 z_0}{\sigma^2}(r^2L\!+\!(3z^2\!-\!r^2)M),\ \
c_{z}^3 \!=\!z^2c_{y}^3,\ \
\nonumber \\&&\!\!\!\!\!\!
c_{z}^4 \! =\!
\frac{y^2z^2}{s^5}(r^2L\!+\!2x^2M).\label{delV1}
\end{eqnarray}

We observe from Eqs.~(\ref{unpert}) and (\ref{delV})-(\ref{delV1}) that $V_i=V_i^0+\delta V_i$ obeys the symmetries,
\begin{eqnarray}&&\!\!\!\!\!\!\!\!
V_x(-x,y,z)\!=\!V_x(x,y,z),\ \  V_x(x,y,-z)\!=\!-V_x(x,y,z), \ \
V_x(x,-y,z)\!=\!V_x(x,y,z),\ \ V_y(-x,y,z)\!=\!-V_y(x,y,z), \nonumber\\&&\!\!\!\!\!\!\!\!
V_y(x,-y,z)\!=\!-V_y(x,y,z),\ \  V_y(x,y,-z)\!=\!-V_y(x,y,z),\ \
V_z(-x,y,z)\!=\!-V_z(x,y,z),\  \  V_z(x,-y,z)\!=\!V_z(x,y,z),\nonumber\\&&\!\!\!\!\!\!\!\! V_z(x,y,-z)\!=\!V_z(x,y,z).\label{symmetries}
\end{eqnarray}
These symmetries, which are rather simple in the case of the infinitely separated wall \cite{ujhd},  are not destroyed by the corrections due to the finiteness of the separation. They allow us to confine the study of the trajectories to the octant $x>0$, $y>0$ and $z>0$.

\end{appendices}


\begin{thebibliography}{99}
\bibitem{mframe} Here, we exploit the fact that the particles travel nearly at a constant velocity relative to the wall despite the relative influence. This allows us to substract the velocity of particle 1, taken as the undisturbed velocity evaluated at its initial center location, from the underlying flow to obtain a moving reference frame. Doing so automatically satisfies the Galilean invariance and significantly reduces the computational cost by limiting the size of the simulation box.
\end{thebibliography}
\end{document}